\documentclass[pageno, 10pt]{jpaper}

\usepackage{authblk}
\usepackage[normalem]{ulem}
\usepackage{amsthm}
\newcommand{\etal}{\emph{et al.\ }}

  \newcommand{\tofix}[1] {
  {\color{red}{NEED-ATTENTION}}}

\theoremstyle{definition}
\newtheorem{defn}{Condition}
\begin{document}

\title{Exploiting Qubit Reuse through Mid-circuit Measurement and Reset}

\author[$\dagger$]{Fei Hua }
\author[$\dagger$]{Yuwei Jin}
\author[$\dagger$]{Yanhao Chen}
\author[$\nabla$]{Suhas Vittal}
\author[*]{Kevin Krsulich}
\author[*]{Lev S. Bishop}
\author[*]{John Lapeyre}
\author[*]{Ali Javadi-Abhari}
\author[$\dagger$]{Eddy Z. Zhang}
\affil[$\dagger$]{Rutgers  University}
\affil[$\nabla$]{Georgia Institute of Technology}
\affil[*]{IBM T. J. Watson Research Center}


 
  \date{}
\maketitle

\thispagestyle{empty}

\begin{abstract}

Quantum measurement is important to quantum computing as it extracts out the outcome of the circuit at the end of the computation.  Previously, all measurements have to be done at the end of the circuit. Otherwise, it will incur significant errors. But it is not the case now. Recently IBM starts supporting dynamic circuit through hardware (instead of software by simulator). With mid-circuit hardware measurement, we can improve circuit efficacy and fidelity from three aspects: (a) reduced qubit usage, (b) reduced swap insertion, and (c) improved fidelity. We demonstrate this using real-world applications Bernstein Verizani on real hardware and show that circuit resource usage can be improved by 60\%, and circuit fidelity can be improved by 15\%. We design a compiler-assisted tool that can find and exploit the tradeoff between qubit reuse, fidelity, gate count, and circuit duration. We also developed a method for identifying whether qubit reuse will be beneficial for a given application. We evaluated our method on a representative set of important applications. We can reduce resource usage by up to 80\% and circuit fidelity by up to 20\%.  

\end{abstract}

\section{Introduction}
\label{sec:intro}

Quantum computation is important as it can solve classical intractable  problems such as factoring \cite{shor:siam99}, chemistry simulation \cite{peruzzo+:naturecomm14,Kandala+:Nature17}, and large database search \cite{grover:astc96}. Due to the spectacular advances in quantum hardware, quantum systems have undertaken significant improvement in the past two decades. Now domain experts can run small-scale experiments on real machines for their specific domain problems.

Quantum measurement is at the very heart of quantum computing. It allows classical systems to extract information from the quantum realm. By allowing repeated executions, measurement can gather information of the final state of a qubit in the form of a discrete probability distribution. Previously on IBM Quantum systems, the measurement is done at the end of a program, for all qubits \cite{tannu+:micro19mea}.

Recently IBM  started providing \emph{hardware} support for \textbf{mid-circuit measurement}, as the very first step for supporting \emph{dynamic circuit} \cite{ibmdc}. It has improved measurement gate duration and fidelity on its Falcon family processors \cite{Govia:arxiv, corcoles+:prl21}. 

Mid-circuit measurement performed when circuit execution is in flight is very useful. It has two types of functionalities: (1) boolean test of state and (2) qubit reuse. For the boolean test of state, it can be used for stabilizer measurement for quantum error correction code \cite{chamberland+:phyrev20} -- to tell if there is an error in the state. It can also be used for asserting certain properties of a qubit for post-processing purposes. Further, it can be used for steering the computation in a useful direction, for instance, the repeat-until-success (RUS) implementation for synthesizing an arbitrary single-qubit rotation gate \cite{paetznick+:arxiv13rus}.   

We focus on the functionality of qubit reuse enabled by mid-circuit measurement. In this case, mid-circuit measurement must be combined with mid-circuit qubit reset, which allows the users to reset the qubit state to the ground state at any point of the program execution. Mid-circuit reset is also supported by IBM hardware recently. After a qubit is reset, it can be reused for any other qubit which hasn't started any operation. This capability of saving qubits is important since today's quantum computer size is in the range of 100 qubits. Being able to efficiently reuse the qubits can increase the capacity of the hardware system to run programs.

\begin{figure*}[htb]
    \centering
    \includegraphics[width=1.0\textwidth]{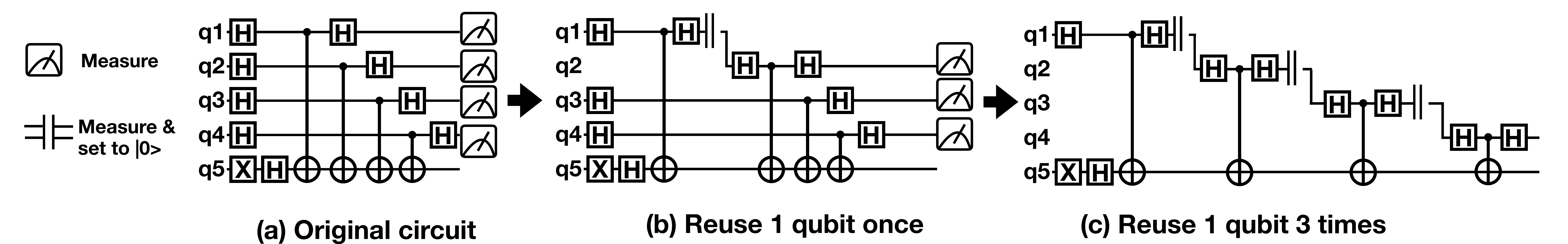}
    \caption{Using Dynamic Circuit Support for the BV Application to Reduce Qubit Usage. (a) Original logical circuit with 5 qubits; (b) Reusing q1 for q2 results in 4 qubits in total; (d) Reusing q1 for q2, q3, and q4 results in 2 qubits in total usage.    }
    \label{fig:bvmot}
\end{figure*}

We show a real application in which qubit count can be significantly reduced by mid-circuit measurement. It is  the Bernstein Vazirani (BV) algorithm in  Fig. \ref{fig:bvmot}. The original circuit uses 5 qubits, but in fact, we can use as few as 2 qubits.  We start with the 1-qubit saving case, where for qubit q1, we can perform measure-and-reset after its last gate Hadamard, as shown in Fig. \ref{fig:bvmot} (b). Immediately we let q1 be reused used for the gates originally applied to qubit q2. We repeat the same process to reuse q1 for q3, and q4 until  we cannot reuse anymore, which results in a 2-qubit circuit, as shown in Fig. \ref{fig:bvmot} (c). In this case, it has reduced as much as  60\% usage of qubit resource. Interestingly, for a $n$-qubit BV application, the minimal number of required qubits is always 2, despite how many qubits are in the original circuit.

Motivated by the resource-saving benefit of qubit reuse, we develop a compiler-assisted approach for automatically transforming circuits exploiting the mid-circuit measurement and reset functionality. Finding proper qubit reuse is challenging. We must also identify the set of applications that can benefit from qubit reuse. With our compiler-assisted tool, users do not have to manually specify whether to reuse or what qubits and when to reuse. Our compiler-assisted tool help users run large quantum programs on small quantum computers. 

Prior studies \cite{tang+:asplos21, Ding+:ISCA20, paler+:physreva16} also optimize qubit usage, but our work is orthogonal to these studies. CutQC \cite{tang+:asplos21} spatially distributes the workload of a quantum circuit  at a cost of worst-case exponential time classical post-processing.  Ancilla qubit reuse \cite{Ding+:ISCA20, paler+:physreva16}  requires un-computation.  The Square framework \cite{Ding+:ISCA20} explores the tradeoff between uncomputation cost and qubit saving. However, our qubit reuse through mid-circuit measurement does not require un-computation and explores a different tradeoff. And in our setting, one qubit can be reused by any other qubit (whether ancilla or non-ancilla) as long as two conditions are satisfied. Therefore the ancilla qubit reuse technique cannot be applied to our problem in this paper.

To our best knowledge, our work is the first to automatically identify qubit-reuse opportunities in general applications. If our tool has identified any qubit-reuse opportunity, it can perform circuit transformation to exploit the trade-off between circuit duration, resource usage, and circuit fidelity. Our tool can handle two different types of applications: the ones with non-commuting gates, and the ones with commuting gates. Furthermore, our tool can also be tuned towards different purposes, towards more qubit saving, or towards gate-count reduction and improved fidelity. 

To summarize, our contributions are as follows:

 \begin{itemize}
     \item We discovered non-trivial potential for qubit reuse in real quantum applications through extensive experimentation.
     \item Our compiler-assisted tool is able to identify if there is any qubit reuse opportunity. If there is any, our tool can yield transformed circuit with respect to different qubit budget level, optimized for circuit duration and fidelity.   
     \item We explore the full spectrum of factors that affect the tradeoff between qubit-reuse and compiled circuit efficiency. We discover that in addition to the benefit of reduced qubit count, qubit-reuse can potentially reduce SWAP insertion and improve circuit fidelity. Hence we designed two versions of our tool such that one emphasizes qubit saving and the other emphasizes SWAP reduction and fidelity.   
      \item Our tool can handle gate commutativity, which is an important feature in modern quantum applications.  
     \item Our experiments show that we can reduce qubit usage by up to 80\%, while keeping circuit duration similar -- slightly larger than the non-reused version by {9.9\%} on average.
     \item We also provide experiment results on real quantum machine IBM Mumbai. We perform experiments on both regular circuit without commuting gates and the applications with commuting gates such as QAOA. In both bases, our results show better performance.  TVD is improved by 17\%. The success rate of finding correct answer increased by 20\%. QAOA can converge faster and find better minimal energy.  Note that they are under the condition of using fewer qubits.  
 \end{itemize} 
 
The rest of the paper is organized as the following. We introduce the background and motivation of qubit reuse in Section \ref{sec:backMot}. We describe the two versions of our compiler-assisted qubit reuse in Section \ref{sec:design}.  We provide  comprehensive experiment evaluation in Section \ref{sec:eval}. Section \ref{sec:rel} describes related work. Section \ref{sec:conclusion} concludes the paper.

\section{Background and Motivation}
\label{sec:backMot}
\subsection{Hardware  Support for Dynamic Circuit}

\begin{figure}[htb]
    \centering
\includegraphics[width=0.45\textwidth]{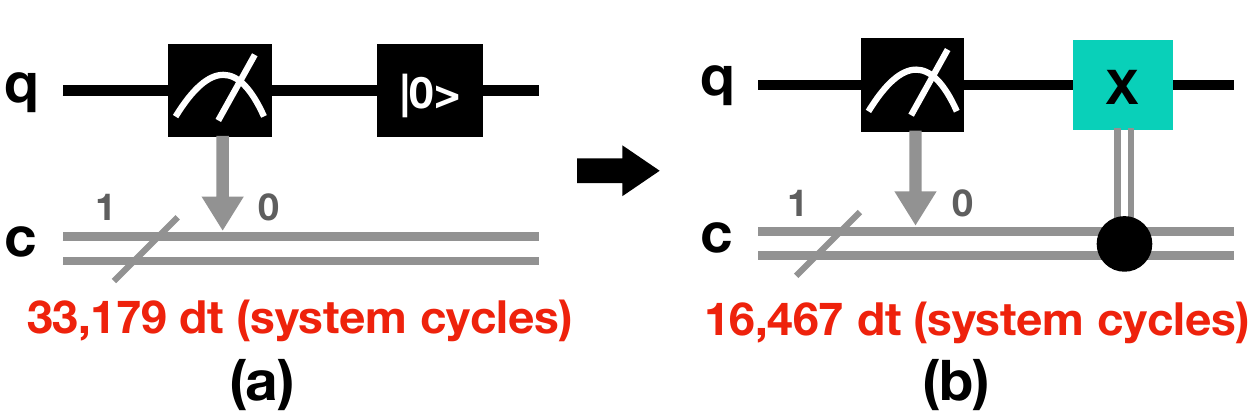}
    \caption{Our improvement for "measurement + reset". (a) Built-in measurement and reset operations in Qiskit; (b) Measurement +  classical control which takes half of the time of (a); } 
    \label{fig:measure_reset}
\end{figure}

 Recently IBM started providing the \textsf{dynamic circuit} support \cite{ibmdc}. In dynamic circuit, it supports  mid-circuit measurement operation and mid-circuit reset operation as the example shown in Fig. \ref{fig:measure_reset} (a). The measurement operation reads out the qubit. The reset operation forces the state of the measured qubit back to the ground state.

 We have made improvements to this combination of measurement and reset. If a qubit is measured as |1$>$ in the standard computation basis, and if we want to reuse this qubit, we must re-initialize it to |0$>$. If a qubit is measured as |0$>$, we do not do anything. However, the built-in reset operation denoted as a box containing |0$>$ implicitly contains measurement pulses, which is redundant. So instead of using a combination of \emph{a measurement + a reset}, we use \emph{a measurement + a classical/quantum control not}. For the classical/quantum control gate, we use the classical bit to control the quantum bit,  as shown in Fig. \ref{fig:measure_reset} (b). We can reduce the duration by around 50\% with this optimization, from  33,179 dt to 16,467 dt, in IBM Mumbai machine. 1dt is 0.22 nano-seconds.

 To improve readability, we use two vertical bars to represent the combination of mid-circuit measurement and conditional reset, as shown Fig. \ref{fig:bvmot} (a) for the rest of the paper.

\subsection{Potential of Qubit Saving}
\label{sec:potential}

\begin{figure}
    \centering
    \includegraphics[width=0.4\textwidth]{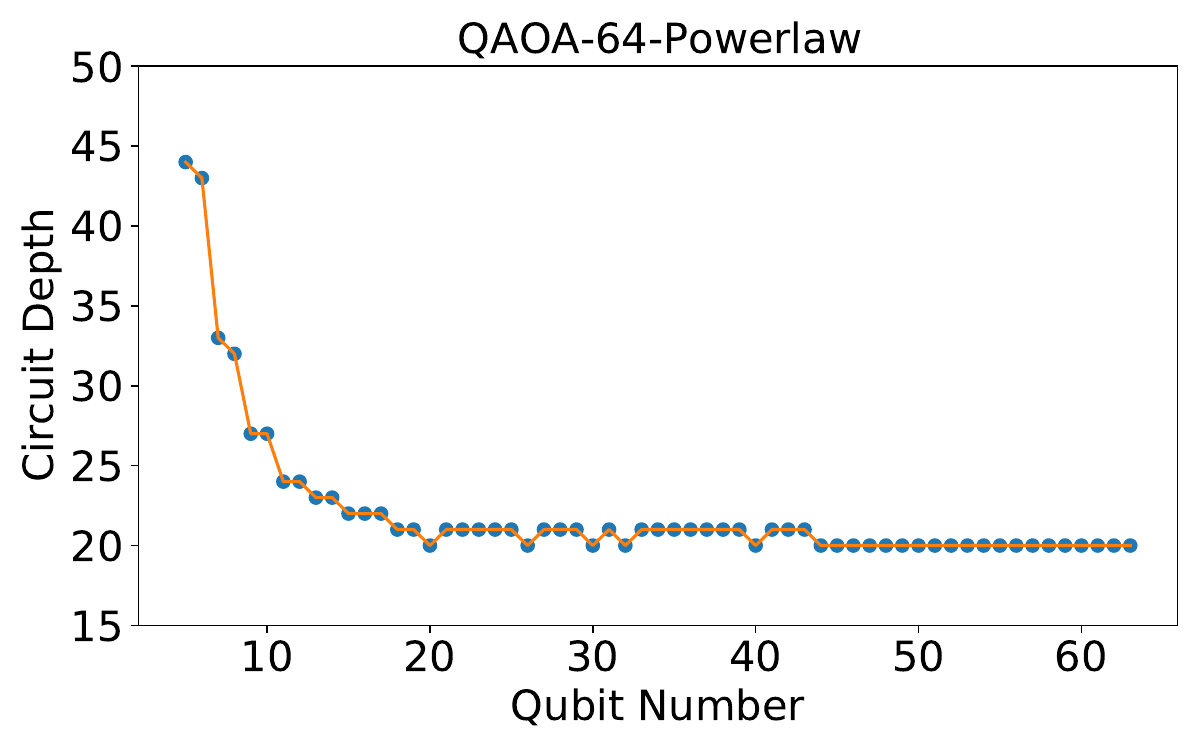}
    
    (a) Input as a power-law graph with density 30\%.
     \includegraphics[width=0.4\textwidth]{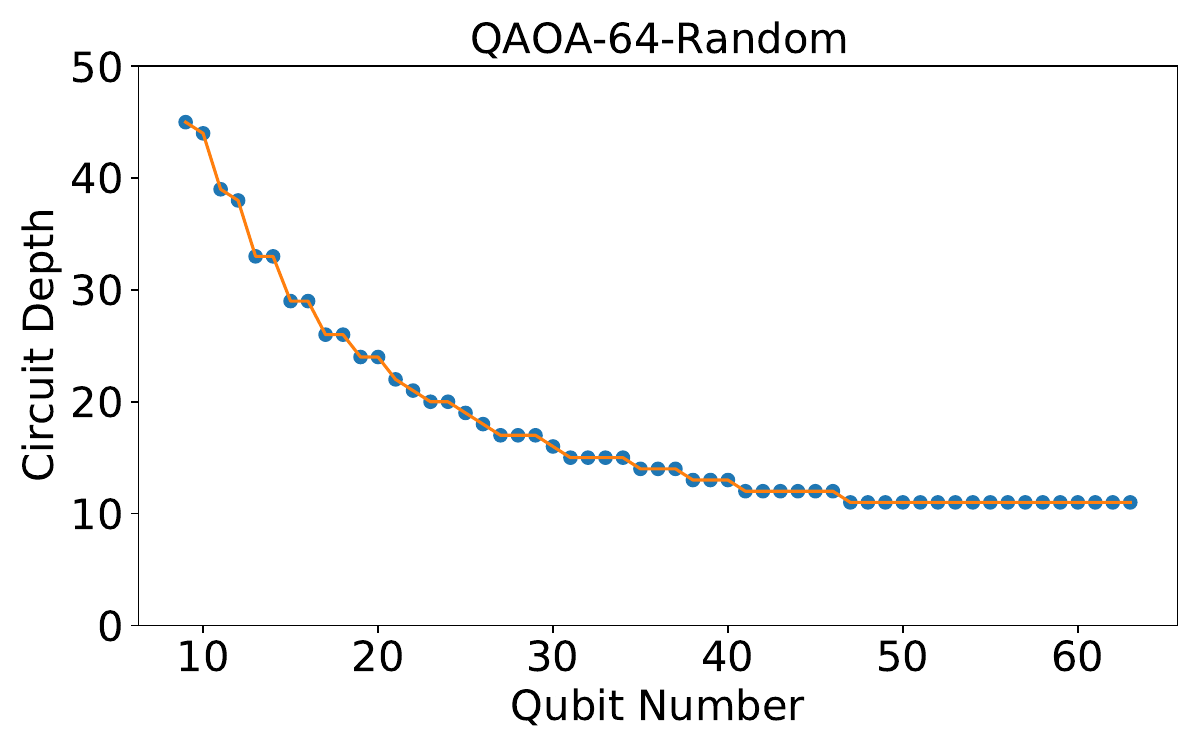}
     
     (b) Input as a random problem with density 30\%. 
      \caption{The Potential of Qubit Saving by Qubit Reuse}
    \label{fig:reusemot}
\end{figure}

To evaluate the potential of qubit reuse for modern quantum applications, we developed a tool that can automatically generate transformed circuit with \emph{(near-)minimal depth/duration} for any qubit reuse count, if such reuse is possible. We have created the data points with all qubit counts. The design of our tool is described in details in Section \ref{sec:qscaqr}. 

We  discovered  that there are significant opportunities for qubit reuse in real applications. We show experiment results of the QAOA application with 64 qubits for two different input problem graphs -- the power law graph and the random graph, both with density of 30\%. It can be seen that qubit usage can be reduced from 64 to as few as 5, which is a significant saving in terms of resources.

It also indicates that as the qubit number decreases, the depth of the circuit increases. However, both cases show (near) heavy-tail distribution, implying that we can potentially reduce qubit usage significantly only by increasing the circuit depth by a reasonably small amount. For the power law graph input  to the QAOA program, we can save over 80\% qubit while only increasing the circuit duration by at most 25\%, as shown in Fig. \ref{fig:reusemot}. For the random graph input, we can save 33\% qubit by increasing the circuit duration by at most 20\%.

\subsection{Tradeoffs for Exploiting Dynamic Circuit}

After demonstrating the large potential of qubit saving by qubit reuse, now we discuss different factors that affect the final circuit performance/fidelity. As aforementioned, increased qubit reuse may increase the depth (or duration if real gate time is available) of a circuit. However, we discover that there are other benefits brought by qubit reuse, which can potentially offset the disadvantage of increased depth. In a nutshell, we can use fewer qubits and in the meantime achieve better transformed circuit's fidelity and performance. 

 We list all benefits of qubit reuse: (1) \textbf{Qubit Saving}, (2) \textbf{Reduced SWAPs}, and  (3) \textbf{Improved Fidelity}, and describe each below.

\paragraph{Qubit Saving} We have already described this benefit of reducing qubit usage in Section \ref{sec:intro} and Section \ref{sec:potential}.

 \begin{figure}[htb]
    \centering
\includegraphics[width=0.4\textwidth]{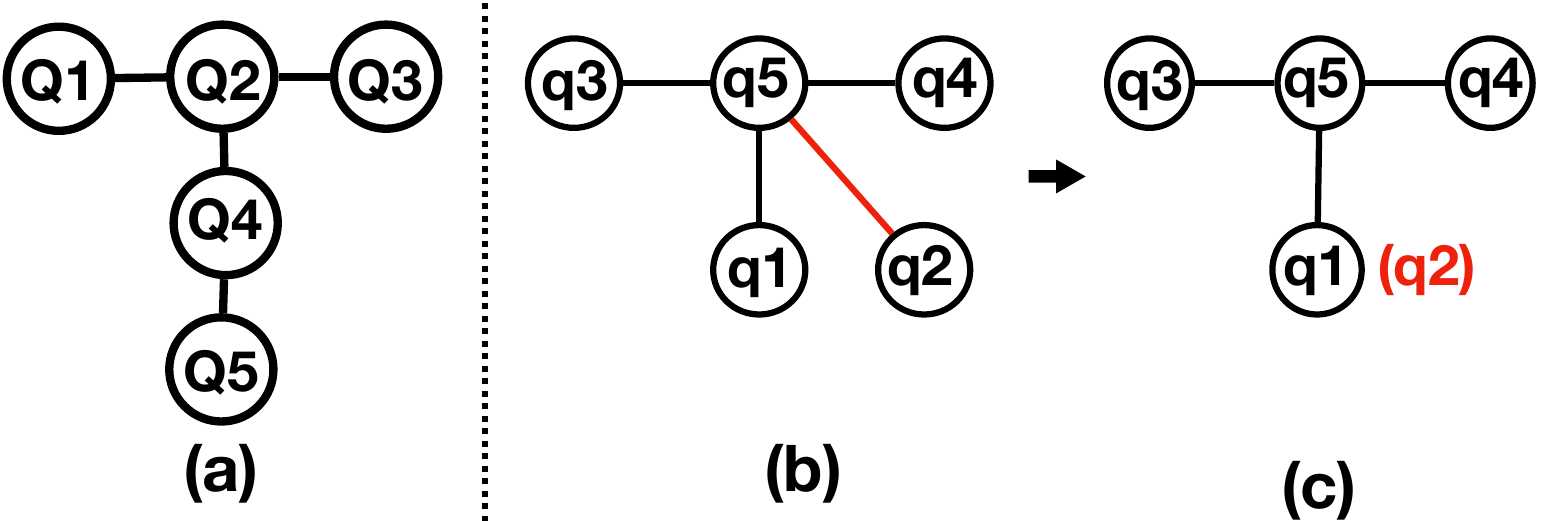}
    \caption{Compiling BV circuit for real architecture. (a) Physical architecture; (b) The qubit interaction graph(if two qubits have a gate, there is an edge) corresponding to the 5-qubit circuit in Fig. \ref{fig:bvmot} (a); (c) The qubit interaction corresponding to the 4-qubit BV circuit in Fig. \ref{fig:bvmot} (b), where q1 is reused by q2. It can be seen that (c) can fit into the physical architecture while (b) cannot. }
    \label{fig:bv_swap}
\end{figure}

\paragraph{SWAP Reduction} In addition to the benefit of reducing qubit usage, we can also reduce the number of swaps through \emph{qubit reuse}. Assuming that the hardware has a 5-qubit physical coupling graph, as shown in Fig. \ref{fig:bv_swap} (a). 

Note the qubit interaction graph for the 5-qubit BV circuit is a star graph, where \emph{q5} has degree of 4 and all other nodes have degree of 1, as shown in Fig. \ref{fig:bv_swap} (b). However, the maximal degree of a physical qubit is 3. Therefore the 5-qubit logical circuit needs to add SWAPs before it is hardware-compliant.

In contrast, the 4-qubit BV circuit with 1 qubit reuse does not need to have any SWAP inserted, if properly mapped to hardware.  The  qubit-interaction graph's maximal degree is 3, as shown in Fig. \ref{fig:bv_swap} (c), by sharing one qubit for q1 and q2. This qubit interaction graph happens to be isomorphic to the hardware architecture in Fig. \ref{fig:bv_swap} (a). No SWAP is needed.

\begin{figure}[htb]
    \centering
\includegraphics[width=0.45\textwidth]{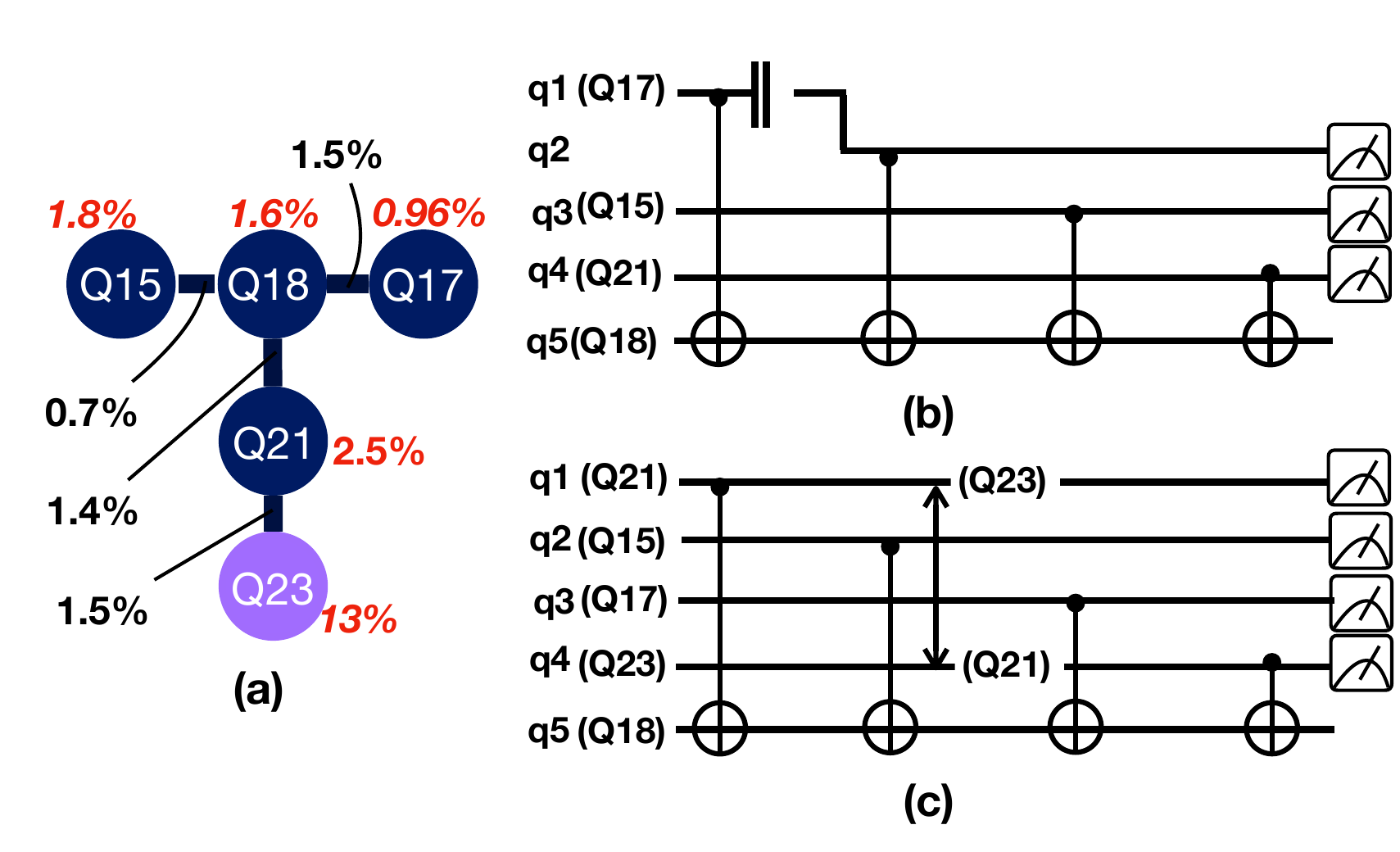}
    \caption{Tackling error variability. For illustration purpose, we eliminate the drawing of Hadamard gates in the figure. }
    \label{fig:bv_diff_q_reuse}
\end{figure}

Even though the 4-qubit logical BV circuit has larger depth than 5-qubit logical BV circuit before SWAP insertion, the final circuit after hardware mapping stage may end up having similar duration because of the additional SWAP(s) inserted. A comparison is shown in Fig. \ref{fig:bv_diff_q_reuse} (b) and (c).  
 
 The reason is that by reusing qubits as if merging multiple nodes into one in the qubit interaction graph, we can potentially alleviate the pressure on hardware connectivity imposed by  the original logical circuit.



\paragraph{Improved Fidelity} 

Qubit reuse happens to have the side benefit of improving the fidelity of the compiled circuit. Today's quantum hardware is presented with the challenge of error variability. Some single-qubit gates or two-qubit gates may have larger error rate than others. By reusing qubits, we can avoid the qubits that particularly have lower fidelity, or the physical links that have higher error rate. 

We still use the same example of BV. We map it to the real IBM machine qubits \emph{Q15, Q17, Q18, Q21, Q23} on Mumbai. The error rates are listed in Fig. \ref{fig:bv_diff_q_reuse} (a). It can be seen that the qubit Q23 has much higher read out error  (13\%) than any other qubit ( $\leq 2.5\%$). By mapping the 4-qubit BV to only qubits of Q15, Q17, Q18, and Q21, as shown in Fig. \ref{fig:bv_diff_q_reuse} (b), we can completely avoid the high readout error on qubit Q23. Since the duration of two circuits are similar as we measured, the overall circuit of 4-qubit BV is actually better. 

\begin{figure}
    \centering
    \includegraphics[width=0.45\textwidth]{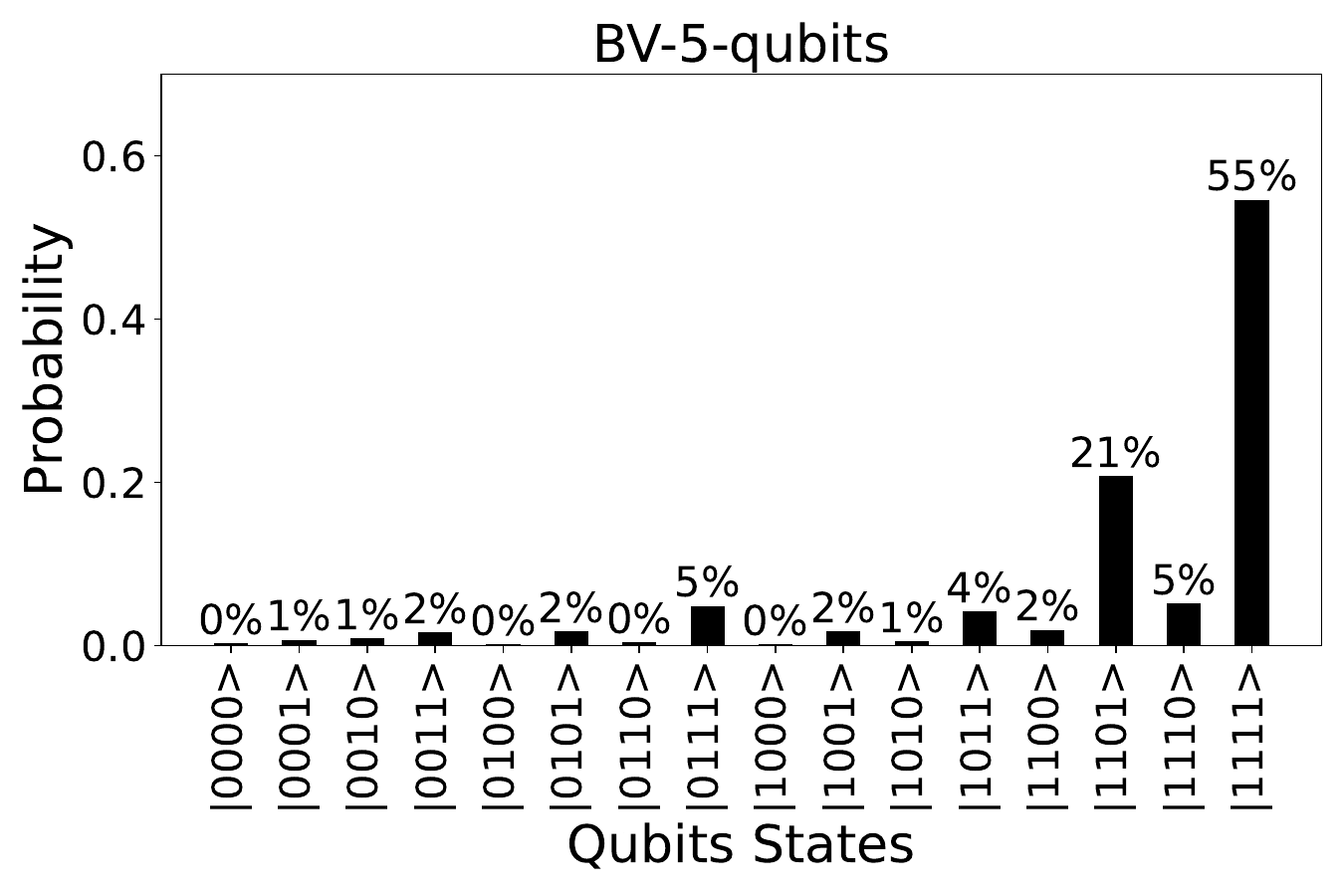}
    {(a) 5-qubit BV on IBM Mumbai on qubits 15, 17, 18, 21, 23}
    \includegraphics[width=0.45\textwidth]{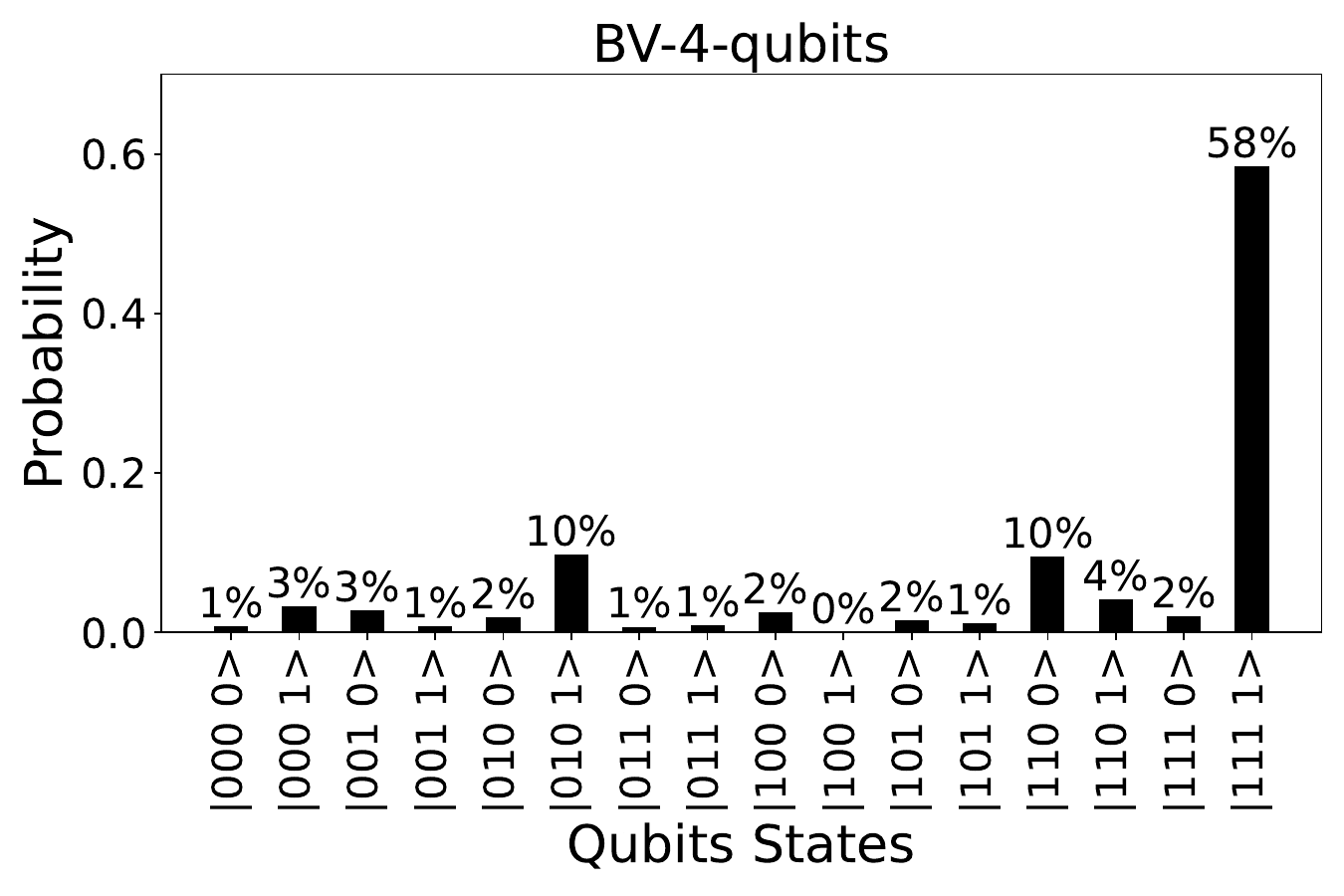}
    (b) 4-qubit BV on IBM Mumbai on qubits 15, 17, 18, 21
    \includegraphics[width=0.45\textwidth]{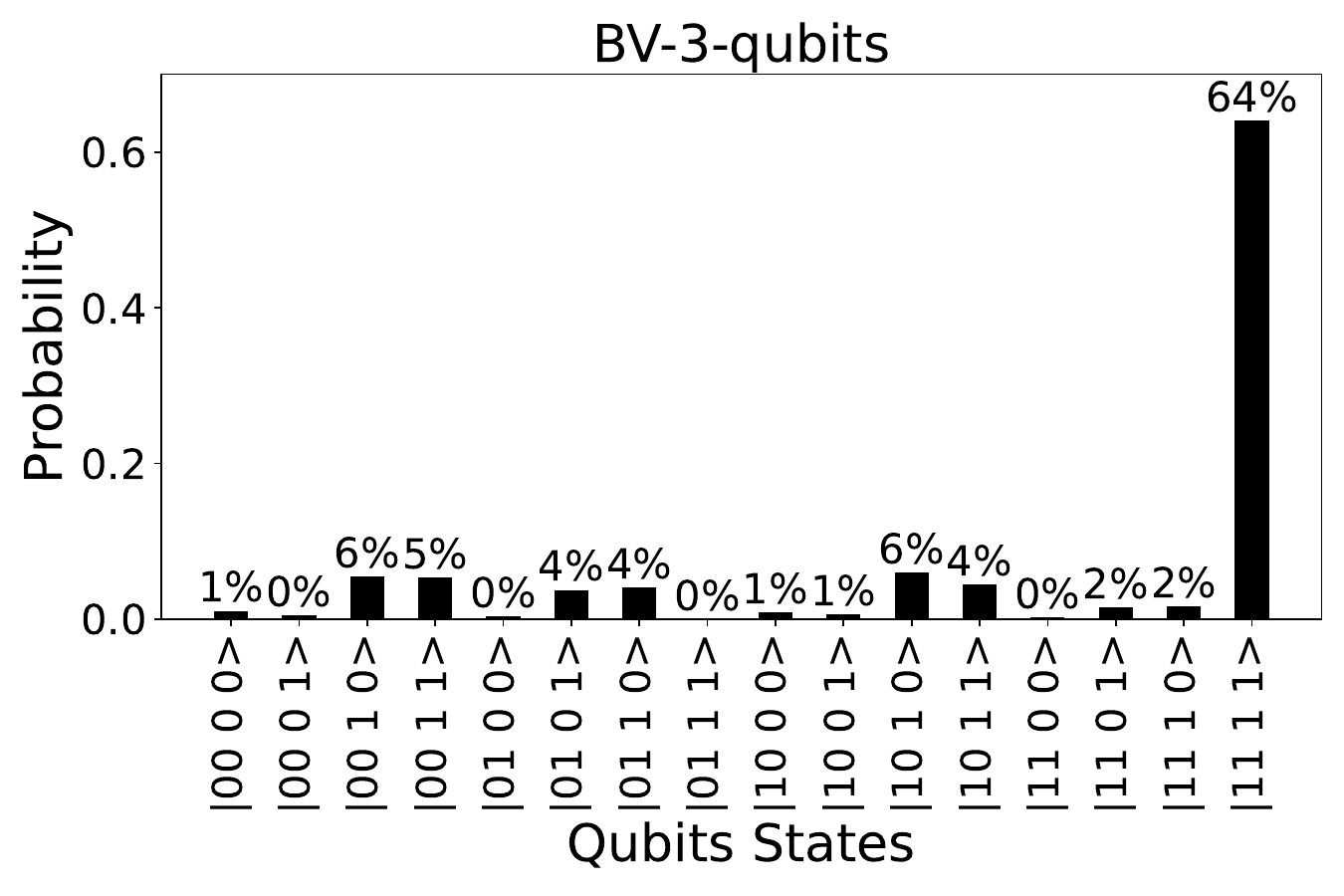}
    (c) 3-qubit BV on IBM Mumbai on qubits 15, 17, 18
    \caption{BV outcome for 5-qubit, 4-qubit, and 3-qubit circuits. }
    \label{fig:real_exp_BV}
\end{figure}

We run this experiment on Mumbai to compare the 5-qubit BV (with a SWAP inserted) with the 4-qubit BV circuit in Fig. \ref{fig:real_exp_BV}. It shows that the 4-qubit BV is better than 5-qubit BV, in terms of the probability of finding the correct answers, 58\% versus 55\%. Note that in addition to higher fidelity, we also saved 20\% qubit usage here. 

Interestingly, we find that using 3 qubits is even better (the circuit duration does not increase significantly due to the dependence relationship in the particular circuit). 3-qubit BV does not use SWAP either. We use Q15, Q17, and Q18, since Q15/17/18 all have better readout error rate than Q21, and the link of Q17-Q18 error is similar to that of Q21-Q18. Now the probability of finding correct answer is improved to 64\%, from 55\% originally, with 40\% qubit saving. In fact, we also tried the 2-qubit BV, but since the circuit duration increase is much larger in 2-qubit BV due to extra measure-reset operations, the 2-qubit BV is not as good as the 3-qubit BV. So we skip the histogram of 2-qubit BV in Fig. \ref{fig:real_exp_BV}.

\paragraph{Discussion} To summarize, reusing qubits with dynamic support is useful in that it can  save resource usage, reduce SWAP gates, and also potentially improving the fidelity. The main disadvantage is increased circuit duration. If the benefits can offset the disadvantage, then qubit-reuse is worth it. It requires a careful analysis to determine which qubits and when to reuse by weighing in all these factors. 


We take these factors into consideration in Section \ref{sec:design}. We provide two version of our design: one can save qubits precisely to user demand, and also allows a tradeoff-based tuning, and the other one primarily improves SWAP insertion.

\section{Design and Implementation}
\label{sec:design}

In our design, we provide two different version of compiler-assisted qubit reuse (CaQR). In the first version, we allow users to precisely control the amount of qubit-saving. We generate a transformed and optimized circuit with respect to a given qubit count, if there is any.  We can also exploit a tradeoff between qubit count, duration (or depth), and fidelity if a range of qubit-saving amount is allowed. We name it as qubit saving CaQR (QS-CaQR). 


To precisely control the number of qubits in a circuit, we must perform qubit-reuse transformation at the logical circuit level. It is because after hardware mapping stage, the inserted SWAPs may reduce the opportunities where we can save qubits. After qubit-reuse transformation is done at logical circuit level, we perform hardware mapping.
 The QS-CaQR version is described in Section \ref{sec:qscaqr}. 

In the second version, we primarily optimize SWAP-reduction through qubit-reuse. We name it as SR-CaQR. For this scenario, resource is not a problem, i.e., there are enough qubits to implement a circuit. But we want to minimize the gate count and in the meantime we want to mitigate the impact of error variability in the architecture.  Hence, we perform dynamic-circuit aware SWAP insertion and save qubit as a side effect. This version is described in Section \ref{sec:srcaqr}. 

Before describing either of the two versions, we first formally define the conditions when a qubit could be reused. 

\subsection{Qubit Reuse Conditions}
\label{sec:reuseConditions}
\begin{defn}
If a logical qubit $q_{i}$ is reused by a logical qubit $q_{j}$, then, there should not be any gate between $q_{i}$ and $q_{j}$. 
\label{defn:reuse1}
\end{defn} 

The Condition \ref{defn:reuse1} for qubit reuse is straightforward. If a logical qubit $q_{i}$ is to be reused by another logical qubit $q_{j}$, we have to make sure all gates on $q_{i}$ finish before all gates on $q_{j}$. If the two qubits have a gate, it is not possible to ensure that. 

\begin{defn}
If a logical qubit $q_{i}$ is reused by a logical qubit $q_{j}$, then, all operations that apply to $q_{i}$ should not depend on any operations on $q_{j}$ directly or indirectly. 
\label{defn:reuse2}
\end{defn} 

\begin{figure}[htb]
    \centering
    \includegraphics[width=0.4
    \textwidth]{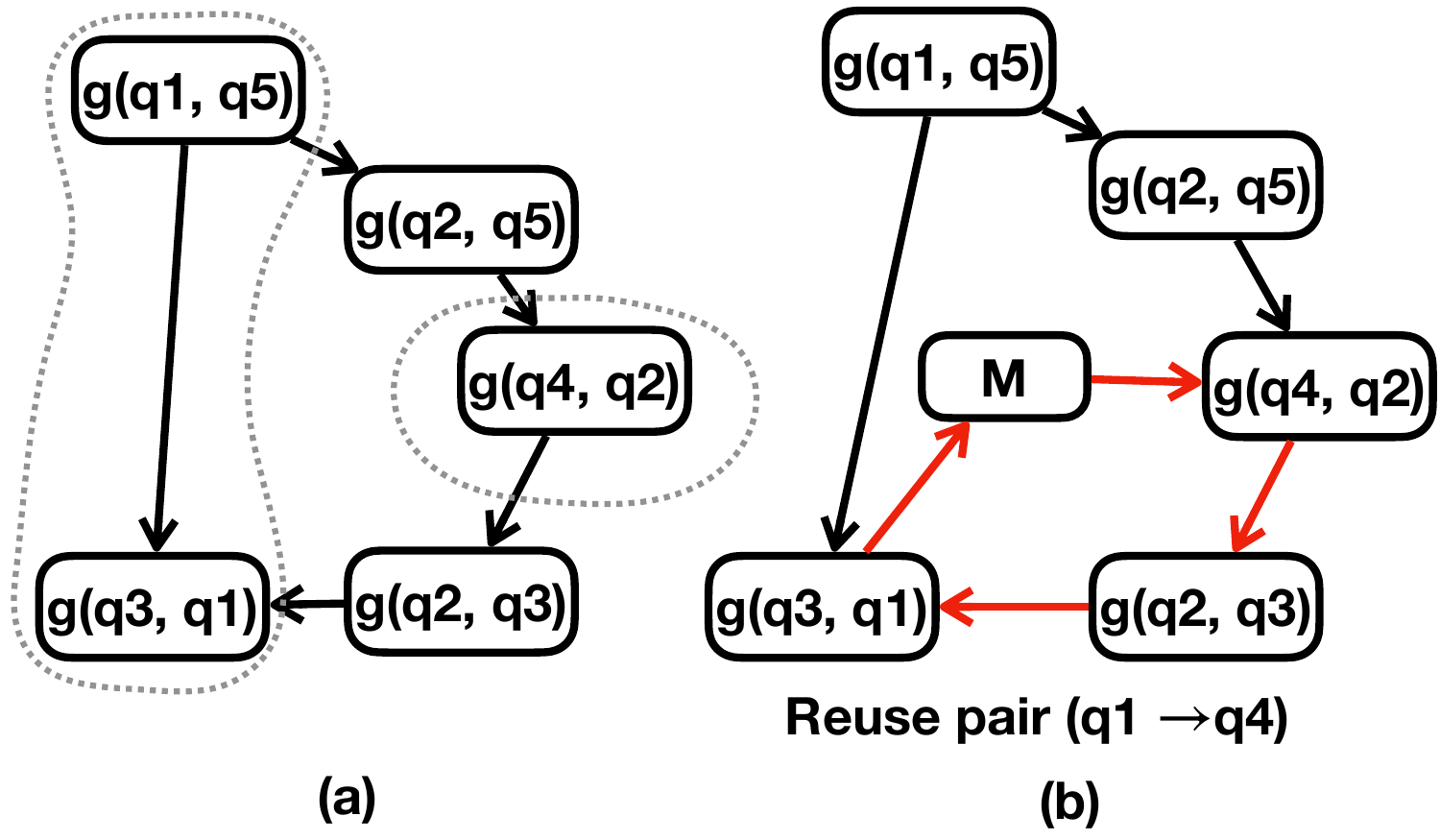}
    \caption{  An invalid qubit reuse pair according to Condition \ref{defn:reuse2}. (a) DAG of the circuit; (b) DAG with measurement-and-reset for the (invalid) qubit reuse pair (q1 $\rightarrow$ q4).}
    \label{fig:condition2}
\end{figure}

{For example, for the DAG graph of a logical circuit in Fig. \ref{fig:condition2}(a), if we reuse q1 for q4 as is shown in Fig. \ref{fig:condition2}(b), gate g(q3, q1) must be finished before gate g(q4, q2), however, gate g(q3, q1) indirectly depends on gate g(q4, q2), leading to a conflict. To detect this automatically, we can see a cycle between the two groups of gates using q1 and q4. And if we insert an M gate standing for the measurement and reuse in between the two groups of gates, the cycle is also manifested. The cycle indicates that a reuse pair is invalid.   }

\subsection{QS-CaQR: Targeting Qubit Saving}
\label{sec:qscaqr}

{\textit{Given a limit of qubits,}}
 we want to see if a circuit can use these limited number of qubits, and if so, provide a compiled and optimized circuit with respect to this qubit count.  
 
 We handle two different types of circuits, one without any commutable gates, and the other with commutable gates. We refer to the former as \emph{regular} applications. 
 
\begin{figure}[htb]
    \centering
    \includegraphics[width=0.3
    \textwidth]{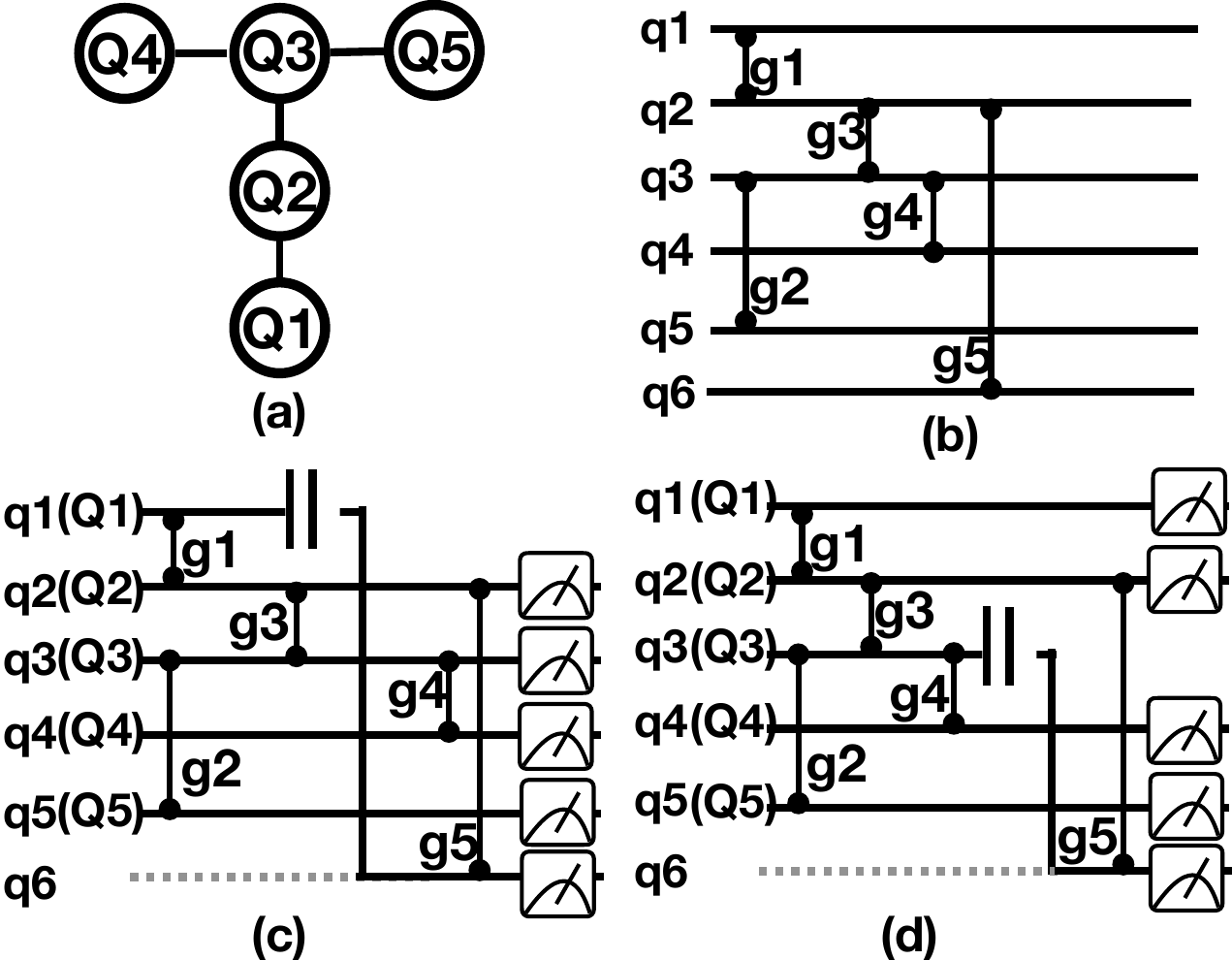}
    \caption{Using different qubits to reuse. (a) Physical architecture; (b) Original  circuit  (c)  Compiled circuit with q1 (Q1) reused. This circuit has depth of 3. (d) Compiled circuit with q3 (Q3) reused. This circuit has depth of 4. }
    \label{fig:strategy_samecount}
\end{figure}

In both cases, we first propose how to find qubit reuse opportunities where the mid-circuit measurement and resetting could apply to. Second, since it's possible that there are many different qubit reuse opportunities, we explore the search space of qubit reuse given the same qubit-limit and choose the best reuse strategy. An example of two different transformations with the same amount of qubit saving but result in different circuit efficiency is in Fig. \ref{fig:strategy_samecount}. 


 \subsubsection{Regular Applications}
 
 To handle a given regular circuit with a limited number of qubits, we first construct a directed acyclic graph (DAG) that encodes the gate dependency. By analyzing the DAG graph, we can check if a qubit could be used by another. We can also analyze the depth of the circuit assuming such reuse takes place. 
 
 We use Condition \ref{defn:reuse1} and \ref{defn:reuse2} to check if a qubit can be reused. With Condition \ref{defn:reuse1}, we check if there is any gate for the reuse pair. If Condition \ref{defn:reuse1} is satisfied, we check Condition \ref{defn:reuse2}.
 
With Condition \ref{defn:reuse2}, we group all the gates that use qubit $q_i$, and then all the gates that use qubit $q_j$. If there are only directed edges (transitively) from $q_i$ to $q_j$, but no directed edges from $q_j$ to $q_i$, then Condition \ref{defn:reuse2} is satisfied. Otherwise, it means there is a cycle if we create such a reuse between $q_i$ and $q_j$, and it means that such a reuse pair is invalid.  

By applying Conditions \ref{defn:reuse1} and \ref{defn:reuse2}, we can find all possible candidate qubit pairs ($q_{i}$ $\rightarrow$ $q_{j}$).  

\begin{figure}[htb]
    \centering
    \includegraphics[width=0.25
    \textwidth]{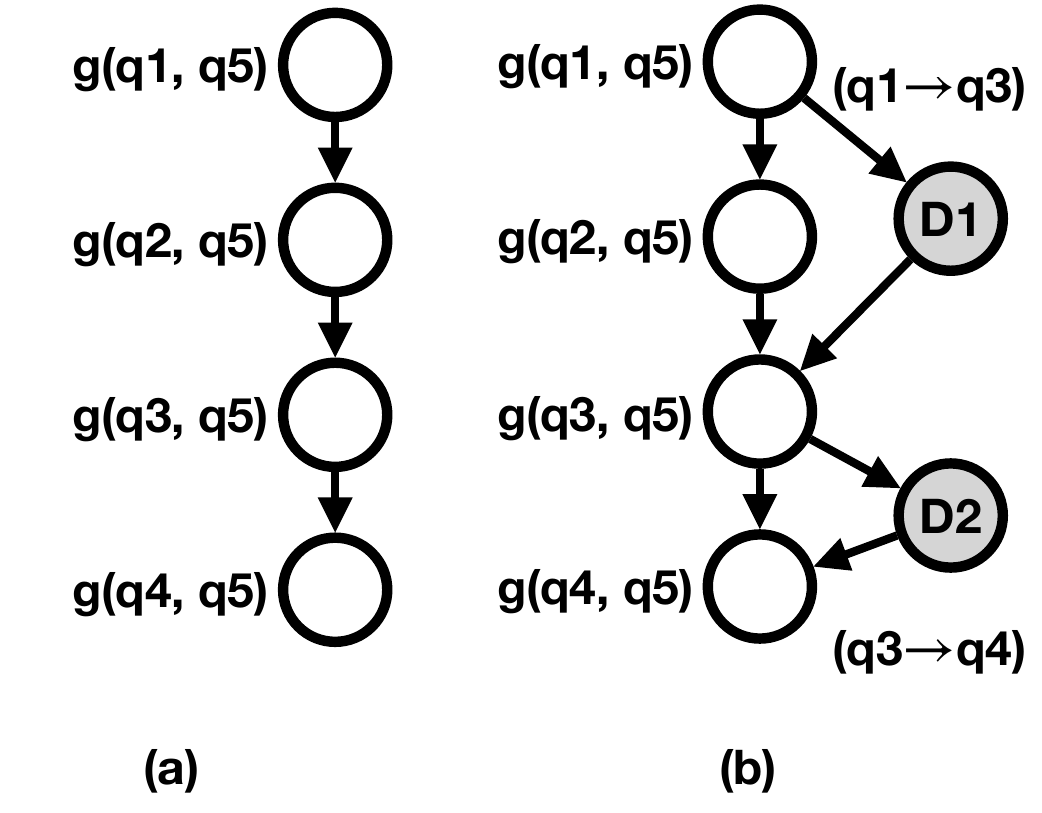}
    \caption{(a) The DAG graph of BV circuit in Fig.\ref{fig:bvmot}(a); (b) Two added dummy nodes in DAG graph for two qubit reuse pairs. }
    \label{fig:regular_reuse}
\end{figure}

 We evaluate one qubit pair at a time. For the qubit pair ($q_{i}$ $\rightarrow$ $q_{j}$), we add a new node $D$ to the DAG graph indicating a measurement-reset operation needs to be applied between the usage of $q_i$ and $q_j$. We make all gates involving $q_{i}$ point to $D$ and making node $D$ point to all gates involving $q_{j}$. 
The qubit reuse pair with smaller critical path length (or circuit duration) in the corresponding DAG is better. For example, {Fig. \ref{fig:regular_reuse}(a) is the DAG graph for BV circuit. To make qubit q1 be reused by qubit q3 first and make qubit q3 be reused by qubit q4 later, we added two dummy nodes D1 and D2 in the DAG graph indicating the new imposed dependency in Fig. \ref{fig:regular_reuse}(b).   }

\paragraph{Qubit Usage v.s. Circuit Duration} Qubit reuse may  potentially increase circuit duration, since the qubit reuse enforces the dependency between two (sets of) gates.  So in the search of qubit reuse opportunity, we must carefully select qubit reuse pair ($q_{i}$ $\rightarrow$ $q_{j}$) and find the one is less harmful to circuit depth.

Our overall strategy is to start with the original qubit count, and gradually reduce it, one at a time, until we reach the qubit count specified by the user. If the user has provided a range of qubit counts, we can generate multiple transformed versions and choose the one with best circuit duration or fidelity (depending on the fidelity metric, for instance, estimated success probability). After logical circuit transformation, we apply a state-of-the-art qubit mapper on it. The candidate circuits for selection are the ones that are finally hardware mapped.

One qubit can be reused multiple times. Our approach allows this type of scenario flexibly, since it picks one qubit-reuse pair at one time. After one pair of reuse pair is picked, we update the circuit, and  keep checking more reuse opportunities.

\subsubsection{Applications with Commutable Gates}

 Another type of circuits is those that have non-trivial amount of gate commutativity. For instance, in the building block of the QAOA application, the CPHASE gates can commute.  So there is no such gate dependency as in regular applications. 
 

 \paragraph{Maximal Qubit Saving} Note that for any given application, there is a bound for the maximal number of qubit saving that can be achieved. For the regular applications, we need to keep reducing qubit count and test if a limit works. But for commuting gates, due to the flexibility of reordering gates, we develop an algorithm that gives the minimal number of qubits. The only constraint for commutable-gates circuit is Condition \ref{defn:reuse1} that two qubits in a reuse pair do not perform 2-qubit gate. This inspired us to use graph coloring to obtain the minimum number of qubits.
 
  We define the qubit interaction graph $G_{int}$ = ($V_{int}$, $E_{int}$), where $V_{int}$ is the set of nodes representing the qubits in the original logical circuit, and $E_{int}$ is the set of edges in $G_{int}$ representing the gates. 
 
 Since the graph coloring algorithm requires that any two connected nodes do not share the same color. So we can apply the graph coloring algorithm on the qubit interaction graph $G_{int}$. The minimum number of color found means the minimum number of qubits needed. For the qubits sharing the same color, one can be reused for another, as long as all operations involving one qubit are finished before any operations involving target-reuse-qubit. 

\begin{figure}[htb]
    \centering
    \includegraphics[width=0.35
    \textwidth]{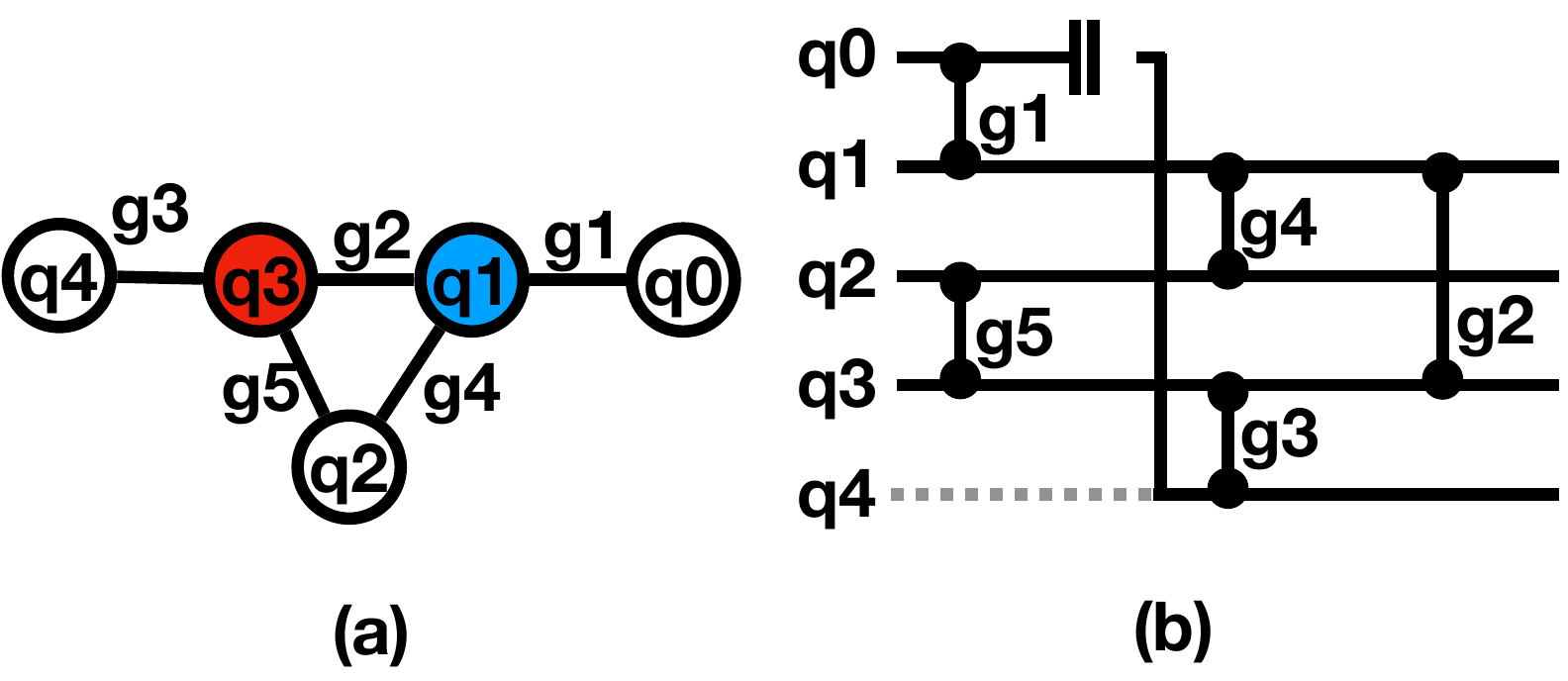}
    \caption{(a) QAOA input graph. Node q0,q2,and q4 in white color. Node q1 in blue and node q3 in red. (b) Transformed QAOA circuit with reuse qubit pair (q0$\rightarrow$q4).}
    \label{fig:qaoa_1q_reuse}
\end{figure}

{For example, we apply graph color to the graph with 5 vertices in Fig. \ref{fig:qaoa_1q_reuse} (a). We found out that the graph  can be colored by at minimal three colors. 
}
 
 \paragraph{Handling Commutativity} To handle a circuit with  commuting 2-qubit gates, we perform it in a similar way to the regular circuits. Firstly, we still list all valid candidate qubit pairs for reuse, \emph{($q_i \rightarrow q_j$)}. Then we evaluate each of them individually. To generate such a candidate set, we have to apply Condition \ref{defn:reuse1} and \ref{defn:reuse2}. Applying Condition \ref{defn:reuse1} is trivial. Two qubits in a reuse pair cannot engage in the same gate. However, Condition \ref{defn:reuse2} is a bit trickier. The original circuit of programs like QAOA does not have pre-determined dependence ordering between gates, hence it is difficult to test if a reuse-pair causes any circle in the execution order. 
 
We handle Condition \ref{defn:reuse2} in the following way. To make the qubit $q_{j}$ reuse the qubit $q_{i}$, we need to schedule all gates on $q_{i}$ first, and gates on $q_{j}$ later. A qubit
reuse pair imposes the gate dependency between two set of
gates. Hence, we let all gates on $q_{j}$ point to all gates on $q_{i}$. As we keep adding reuse pairs, there are more and more dependence edges added to the graph. Each time we add a qubit reuse-pair, we can test if it creates a cycle in the dependence graph with respect to Condition \ref{defn:reuse2}. If not, then such a reuse pair is a valid pair. 

\paragraph{Algorithm for Evaluating the Impact of A Reuse} Now we describe our algorithm for evaluating the impact by applying a particular qubit-reuse pair. We maintain a list of gates in the frontier, that is, the set of gates that either do not depend on any other gate due to qubit-reuse or its dependence is resolved. Every iteration, we choose gates in the frontier to schedule, and we repeat this until no gates are left. That is, until no edges are left in $G_{int}$. Below are three steps:

\begin{figure}[htb]
    \centering
    \includegraphics[width=0.45
    \textwidth]{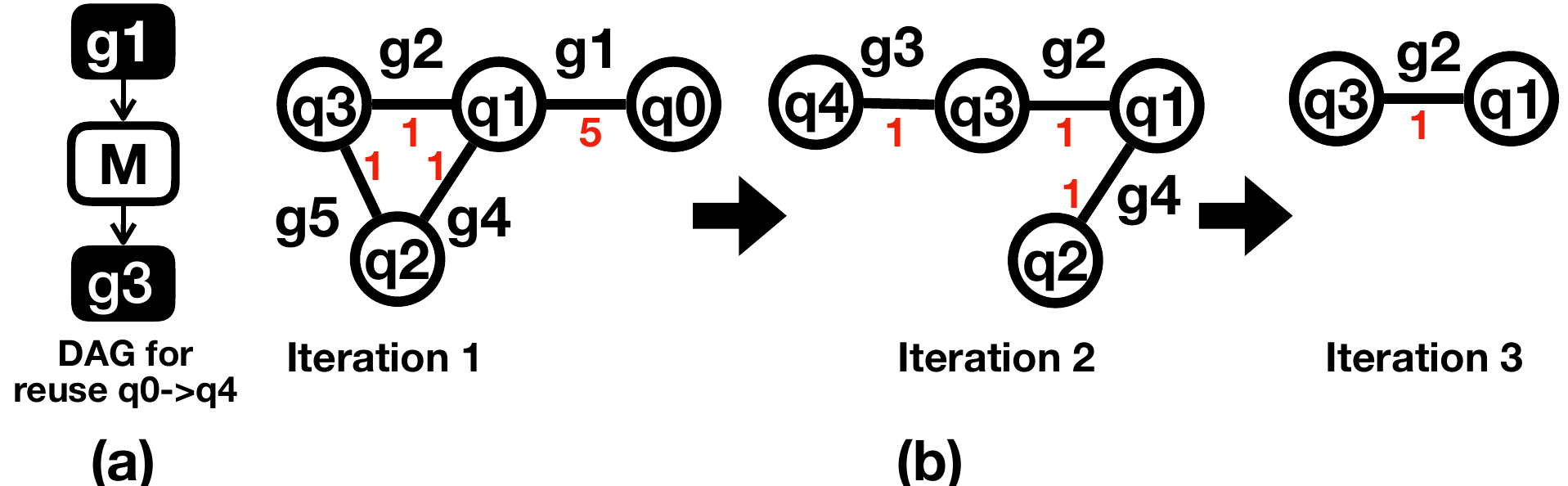}
    \caption{(a) Paitial DAG graph for QAOA circuit corresponding to Fig. \ref{fig:qaoa_1q_reuse} (a) with reuse pair (q0$\rightarrow$ q4). The M in the middle stands for a measurement and reset; (b) Gates g1 and g5 are selected at iteration 1 after perfect matching; Gates g3 and g4 are selected at iteration 2; and gate g2 is selected at iteration 3. The weight of edge is in red. Transformed circuit in Fig. \ref{fig:qaoa_1q_reuse}}
    \label{fig:qaoa_1q_detail}
\end{figure}

\textbf{Step 1}: We use the pair ($q_{i}$ $\rightarrow$ $q_{j}$) to update the current dependence graph $G_{D}$, or to create dependence edges if it is the first qubit-reuse pair to apply.   The qubit reuse pair imposes the dependency on the gates related to $q_{i}$ and $q_{j}$. We add a new node that represents the  measurement-and-reset operation between the two sets of gates. That is, all gates involving $q_{i}$ should point to the new gate, and all gates involving $q_{j}$ should be pointed to by this new gate. 

\textbf{Step 2}: We temporarily remove all gates that correspond to the gates with none-zero in-degree in the dependence graph $G_{D}$ from the qubit interaction graph $G_{int}$. It is because those gates depend on other gates and cannot be scheduled before the dependency resolved. 

Now we assign weights to the edges in $G_{int}$.  We want to prioritize the gates that has gates depending on it -- the type of gates involving qubits to be reused.  So we assign larger weights to these gates as a parameter $|E_{int}| > 1$, while all other edges have weight 1. 

\textbf{Step 3}: We apply maximum weight perfect matching algorithm to $G_{int}$. Those gates with higher priority would be more likely selected. The maximal matching algorithm will select as many gates as possible to improve the parallelism. Those selected gates are scheduled and the corresponding edges in $G_{int}$ and nodes in $G_{D}$ are removed permanently. Then we place the temporarily removed gates back to $G_{int}$. We now go back to \textbf{Step 2} until the frontier has no gate. 

With the above three steps, we can get the duration of the circuit after applying the reuse pair. Since we have tried to maximize the parallelism by perfect matching, we got a circuit that has  good duration. We associate the duration with such a qubit reuse pair. {A full example for how we transform the QAOA circuit  is shown in Fig. \ref{fig:qaoa_1q_detail}.}

 Similarly, for this type of applications, our compiler takes a circuit as input, and a limit of qubits.  It outputs two types of results: (1) Yes, there exists a way to build the circuit with this limit on qubits, or (2) no.
For each given qubit limit, we process it in the same way as that for regular applications. We evaluate and rank different qubit-reuse pairs, and choose the best one with respect to circuit duration/fidelity. At each time, we save one qubit. Then we keep saving until we reach the given qubit limit, or until no more qubit can be saved. If we are given a range of qubit-counts, we will be able to generate different versions of logical circuits that can be used for further selection with respect to user requirement.

\subsection{SR-CaQR: Targeting Reduction of SWAPs and Improved Fidelity}
\label{sec:srcaqr}

Now assuming we have enough resources. We design SR-CaQR to compile a given circuit and treat the SWAP gate reduction as the primary goal through qubit reuse. 

The main reason we can achieve SWAP gate reduction is because we can delay the first gate for some qubits without extending the critical path. Those qubits that have not started any operations can map to two type of physical qubits. One is the fresh physical qubit that is not used by any logical qubits. The another one is the used physical qubit but which has done all operations on that. With a broader selection of physical qubits, we can pick the one (or two) with smaller distance (or with lower error rate) in architecture coupling when one (or two) new logical qubit has to be mapped such that the SWAP gates are reduced. Since the gate delay would not extend the critical path of circuit but SWAP gates are reduced, the depth of compiled could be potentially reduced as well.  

SR-CaQR considers the solution for both application types as QS-CaQR. The details of the design are in following.

\subsubsection{Regular Applications} 
For regular program compilation, we take the logical circuit and the hardware information as the inputs. Then, we compile the circuit layer by layer and map the logical qubit to physical quit when necessary. The new logical qubit will pick the best available physical qubit to minimize the SWAP gates snd improve fidelity. 

\textbf{Step1}: We construct the DAG graph $G_D$ for the input circuit and maintain a list of physical qubits that are available to use in the list $physicalList$. Initially, $physicalList$ contains all physical qubits. By doing the analysis on DAG graph, we can easily find out that whether a gate is in critical path or not. 

\textbf{Step 2}: Considering the gate with qubit(s) not mapped. For all gates with \emph{in-degree = 0} in the frontier of DAG, if the gate is not on the critical path, we delay it. If the gate is on the critical path, we must run it. It it possible that this gate has both qubits not mapped, then we map the qubit with more gates on it first. This logical qubit will pick a physical qubit from $physicalList$ that also can benefit future gates involving it (by lookahead). Or it will map to a physical qubit with better connectivity. The mapped physical qubit is then removed from $physicalList$.

For another unmapped logical qubit of the gate with only one qubit already mapped, we pick the one from $physicalList$ with minimum distance to its already mapped qubit. If there is a tie, we pick the qubit with smaller readout error or the qubit connected by a physical link with smaller CNOT error. The mapped physical qubit is  removed from $physicalList$.

\textbf{Step 3}: Considering the gate with both qubits mapped. For all gates in the frontier of DAG whose both qubits are mapped, if the gate is hardware-compliant, we schedule it. If two qubits are not adjacent to each other, we add SWAP gates. We use a heuristic method to insert SWAP gate with the consideration of error variability and the side-effect on the following gates. 

\textbf{Step 4}: If any gates are scheduled, we update the frontier in DAG graph $G_D$. If there is any qubit done with all operations on it, we added this qubit back to $physicalList$. We repeat the step 2-4 until the frontier is empty.

\begin{figure}[htb]
    \centering
    \includegraphics[width=0.35
    \textwidth]{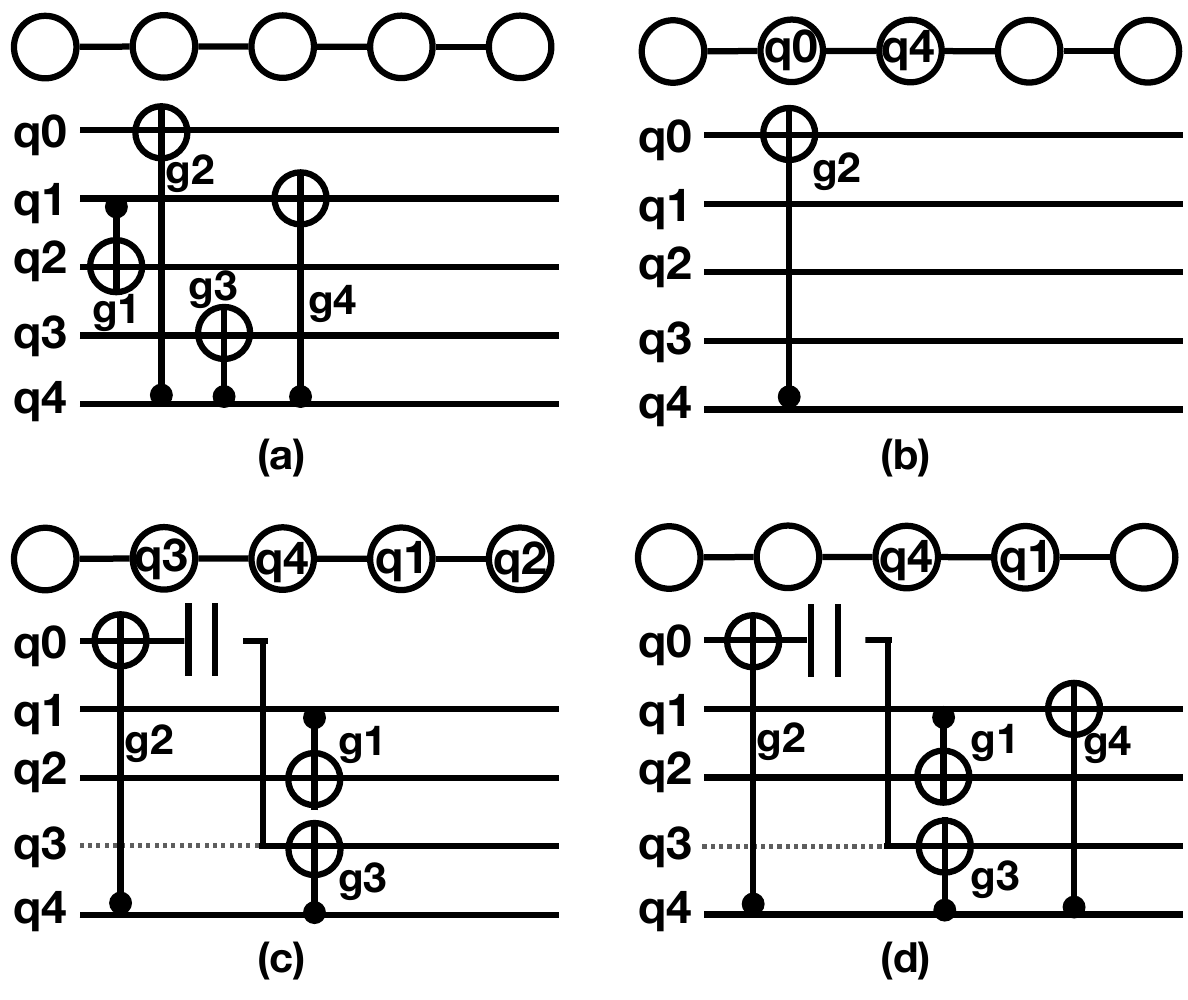}
    \caption{Example of SR-CaQR for regular applications}
    \label{fig:SR_CaQR_regular}
\end{figure}

We use an example to explain our method.  Fig. \ref{fig:SR_CaQR_regular} (a) is a logical circuit with the physical coupling graph on the top of it. There is nothing mapped on the coupling graph. In Fig. \ref{fig:SR_CaQR_regular} (b), only two qubits q4 and q0 are mapped and g2 are scheduled. This is because gate g1 is not on the critical path which is delayed and g2 is on the critical path. q4 is mapped first since more gates apply on it and is mapped to the middle of the architecture coupling graph. 

In Fig. \ref{fig:SR_CaQR_regular} (c), after gate g2 is scheduled, we reclaim the qubit q0 and save it to $physicalList$ since no more gates on qubit q0. Now gate g1 and g3 are both on the critial path. For gate g1, we map q1 first since it involves more gates. Qubit q1 will interact with q4 later, so we map it close to q4.  For gate g3, q4 is already mapped, then we map q3 close to it. 

In Fig. \ref{fig:SR_CaQR_regular} (d), we free q3 and q2 since they are done with all operations. Gate g4 has two mapped qubits and is hardware-compliant. 

In this example, SR-CaQR added zero SWAP gate by applying qubit reuse. However, the original circuit at least takes one SWAP gate to make the circuit hardware-compliant.

\subsubsection{Applications with Commuting Gates}

SR-CaQR also considers the solution for compiling applications with commutable gates such as QAOA. The main idea is similar to that for regular applications. We try to delay the start time of some qubits such that those qubit would have more options of candidate physical qubits. However, unlike the regular applications, applications with commutable gates do not have gate dependency. To solve this problem, we can utilize QS-CaQR to manually impose gate dependency for a part of the gates. The details are in the following. 

\textbf{Step 1}: Constructing a parital DAG $G_D$. For the given QAOA circuit, we use QS-CaQR to find a sweet point of number of qubit reuse and the corresponding qubit reuse pairs \emph{($q_i \rightarrow q_j$)}. Then, we construct a partial DAG based on the qubit reuse pairs. All gates that apply to qubit $q_j$ depend on all gates that to qubit $q_i$. The DAG $G_D$ also contains gates that do not involve the qubit reuse. Those gates have in-dgree = 0. After graph $G_D$ constructed, we update the degree information since the imposed qubit reuse and gate dependency increase the degree of $q_i$. Then, we start scheduling those gates with \emph{in-degree = 0}.

\textbf{Step 2}: Delaying gates. For the gate with one qubit or two qubits not mapped, we want to delay it based on the following two conditions. Firstly, if the gate is in the reuse dependency graph, we do not delay it. Since all gates in the reuse dependency have to be scheduled first such that the corresponding qubit can be measured and reused by other reuse-target qubits. Secondly, if the gate has one or two qubits with high degree on QAOA graph, we do not delay it. The highest degree in the QAOA graph determines the lower bound of depth of compiled circuit, so delaying gate with high degree qubit would potentially increase the circuit depth.

\textbf{Step 3}: Scheduling gates. If the gate is not delayed or has both qubit  mapped, we schedule it. For the gate with both qubits  mapped, we schedule it if it is hardware-compliant. Otherwise, we add SWAP gates for it heuristically. For the gate with one or two qubits unmapped (but not delayed), we map its logic qubits with the consideration of qubit distance and error variability which is the same as the regular application solution. Then, we schedule it or add SWAP gates for it. 

\textbf{Step 4}: Update information and repeat. We remove those scheduled gates from $G_D$ and update the degree information. Here, we still reclaim qubits if they are done with all their operations.  
Then we repeat step 2-4 until $G_D$ is empty.


\section{Evaluation}
\label{sec:eval}
In this section, we evaluate our proposed methods by using different types of quantum circuits: regular circuits without commutable gates, and circuits with commutable gates. We explore qubit reuse opportunities in a given circuit and analyze the trade-off between qubit usage, circuit depth/duration, and gate count. We also perform real machine experiments to show that qubit reuse could help improve end-to-end circuit fidelity and performance.

\subsection{Experiment setup}
\paragraph{Architectures and Backend}  Both QS-CaQR and SR-CaQR are using IBM heavy-hex as the backends. When the qubit number is large, we use the scaled heavy-hex architecture. 

For real machine, we use IBM Mumbai which also exhibits heavy-hex pattern. We also use the real calibration data exported from the IBM systems including the CNOT duration, CNOT error for each physical link, and qubit readout errors. 

\paragraph{Metrics} To assess our proposed method, we use {qubit usage},  {two-qubit gate count}, and {circuit depth/duration} as metrics. We use total variant distance (TVD) if necessary. We also use the final application outcome on real machines as a way to evaluate the effectiveness of our method.  

 Qubit reuse enforces extra gate dependency in the circuit and potentially increases the circuit duration. A circuit with a smaller duration has less decoherence error. So we need to evaluate circuit duration. Two-qubit gate count is another concern. Qubit reuse could potentially save the number of SWAP gates inserted. Since we can reuse a nearby qubit if both conditions in Section \ref{sec:reuseConditions} are satisfied to reduce the communication cost for a gate involving two distant qubits.

\paragraph{Baselines} We use IBM Qiskit as the baseline, with optimization level 3 turned on. It compile both regular and commutable-gate circuit. QAOA is classical-quantum application. Our optimization is on the quantum part. It also needs a classical optimizer. For running QAOA for the full experiment, we use well-known "COBYLA" classical optimizer provided by IBM Qiskit by default.

\paragraph{Benchmarks}
We have two types of benchmarks, regular quantum applications, and commutable-gate quantum applications.  We have the regular quantum applications: Rd-32, 4mod5, Multiply\_13, System\_19, CC\_10, XOR\_5, and BV\_10 \cite{cross+TOQC2022}\cite{li+2021qasmbench}. For the commutable-gate circuits, we use QAOA circuits for max-cut problem. We have two different types of input problem graphs, random graph and power-law graph, with different sizes, from 16 to 128.

\subsection{QS-CaQR Evaluation} 
In this section, we evaluate the qubit saving version: QS-CaQR. QS-CaQR first  applies qubit reuse on logical circuits. 

For each application, we tried different qubit limit numbers, and generate different compiled circuits. 
For each circuit corresponding to a desired qubit usage, we use Qiskit transpiler to insert SWAPs. We show the results for both regular applications and QAOA applications.

\subsubsection{Regular Applications}

\begin{figure}[htbp]
     \centering
     \subfloat[]{\includegraphics[width=0.5\textwidth]{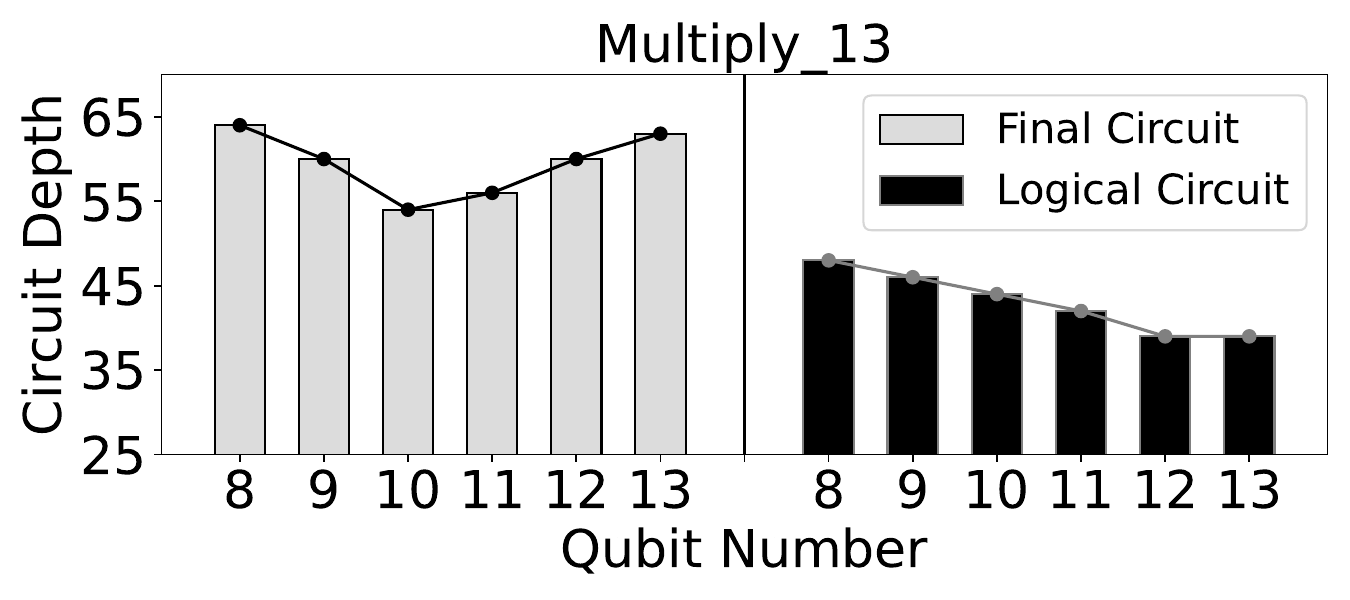}}

    \subfloat[]{\includegraphics[width=0.5\textwidth]{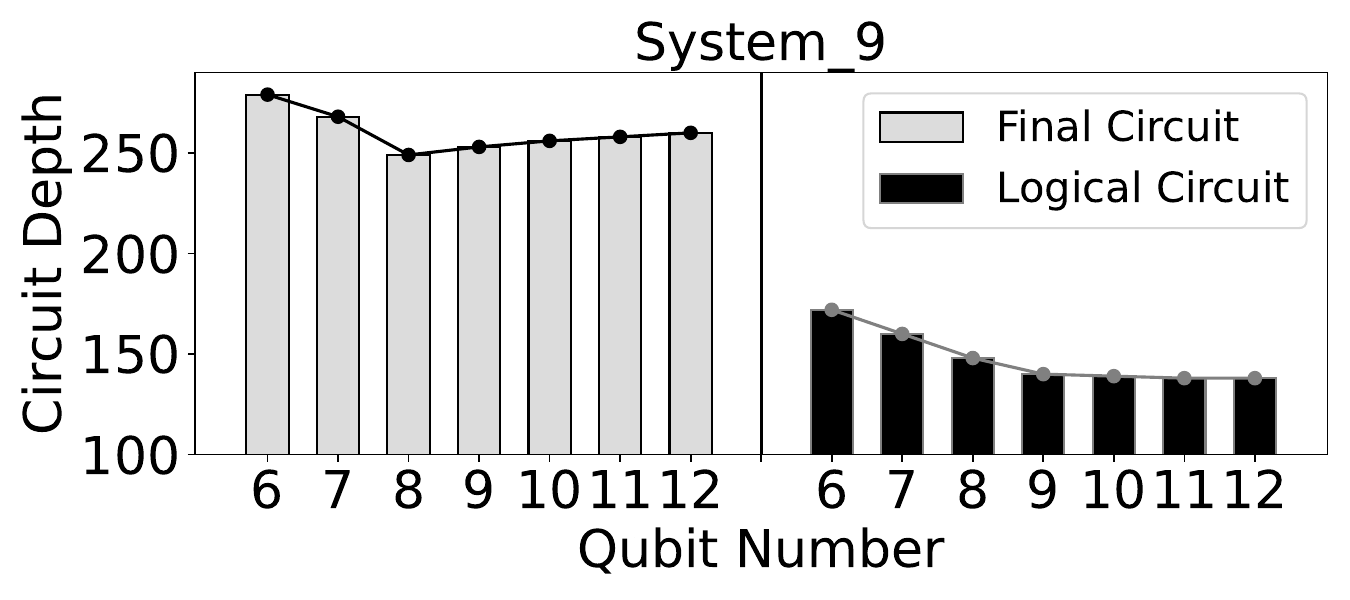}}

     \subfloat[]{\includegraphics[width=0.5\textwidth]{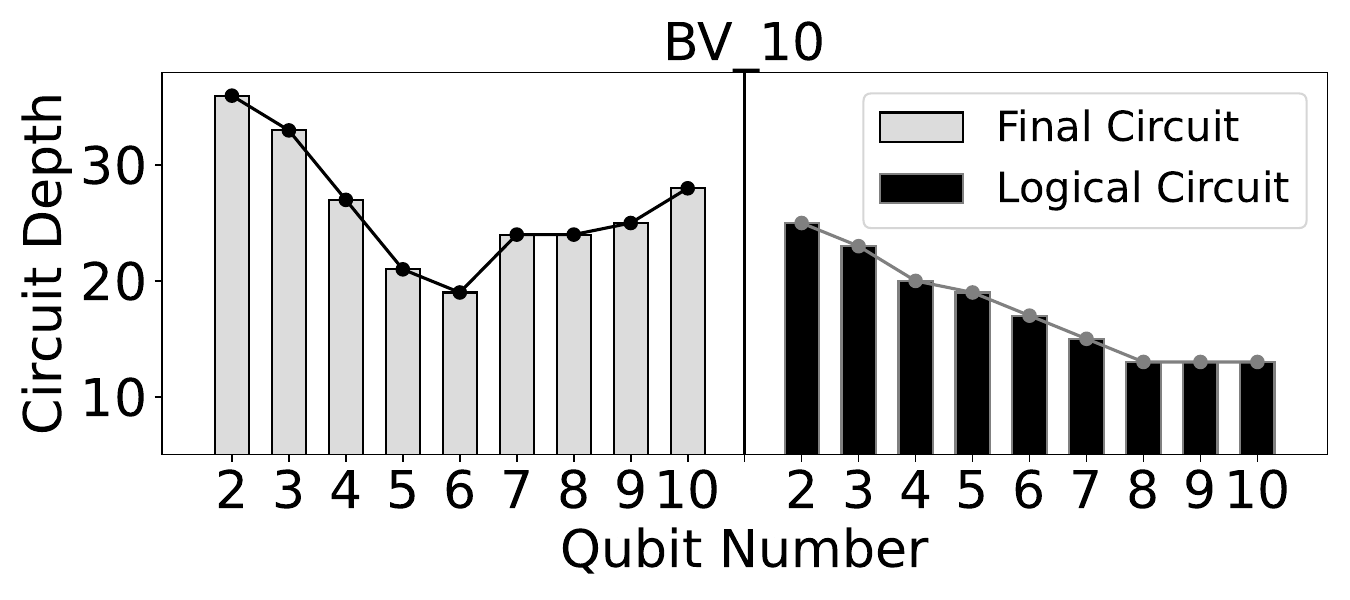}}

     \centering

        \caption{QS-CaQR: Reuse vs depth for regular  circuits. }
        \label{fig:nonQAOA_noSWAP}
\end{figure}

We show the results of three regular applications, Multiply\_13, System\_9, and BV\_10 in Fig. \ref{fig:nonQAOA_noSWAP}. The bars on the right half of Fig. \ref{fig:nonQAOA_noSWAP} (a), (b), and (c) represent the depth of logical circuits with respect to qubit usage reduction, and the grey bar plots on the left half figures represent the depth of final compiled circuits with hardware mapping. We can see that, for all logical circuits, when the number of qubit usage decreases (from right to left in the x-axis), the circuit depth increases. However, the results for the final circuit show a different pattern. When qubit number decreases, the depth decreases slowly in the beginning and increases in the end. It is potentially because when qubit-saving is too aggressive, it does not help SWAP insertion. 

Depending on user request, we can choose different compiled circuits. If the users demand to have a minimal depth circuit, we need to choose a version that saves some qubits moderately. The sweet spot is usually in the middle. If the users demand to save qubits, it will not have best depth.

\subsubsection{QAOA Applications}

\begin{figure}[htbp]
     \centering \subfloat[]{
 \includegraphics[width=0.45\textwidth]{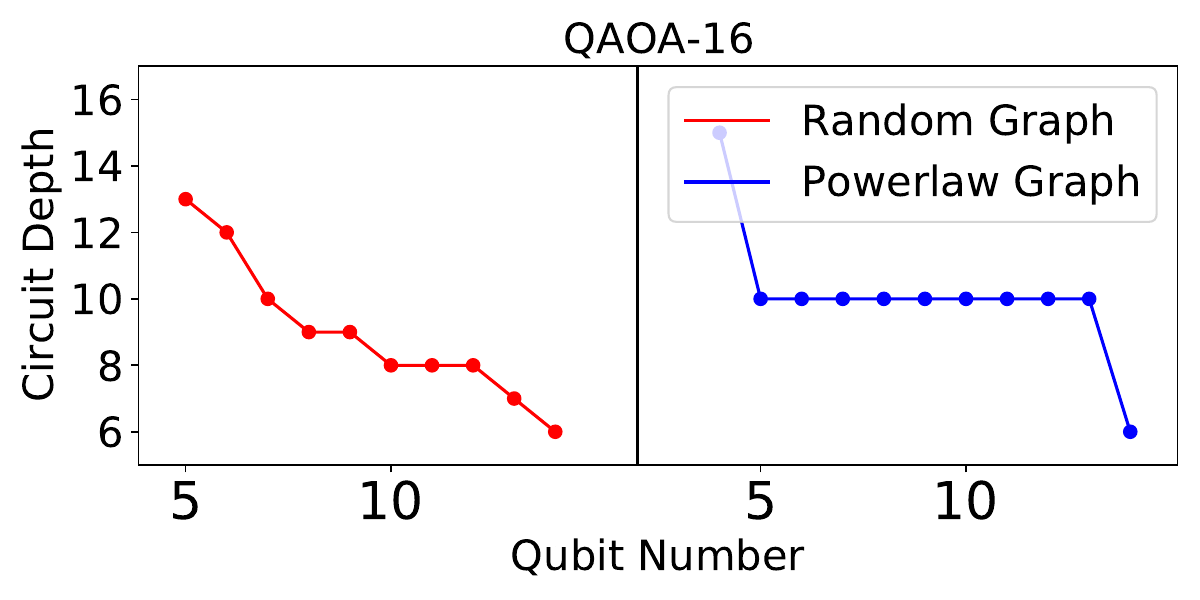}
         }
         \vspace*{-1.0em}
         \subfloat[]{
 \includegraphics[width=0.45\textwidth]{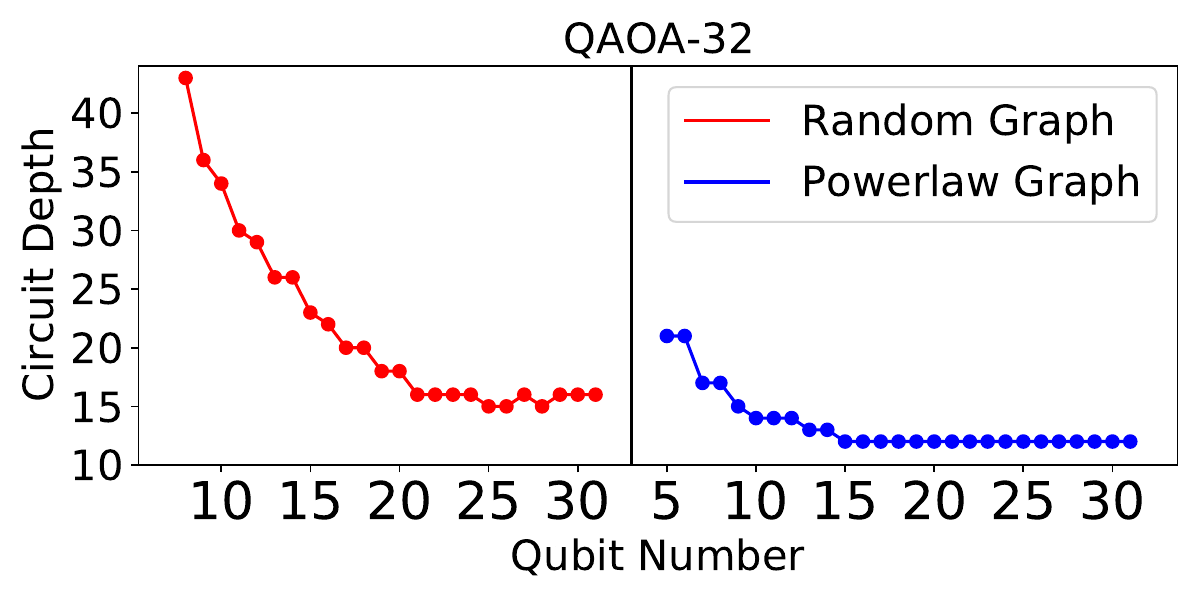}
         }
         
         \subfloat[]{
 \includegraphics[width=0.45\textwidth]{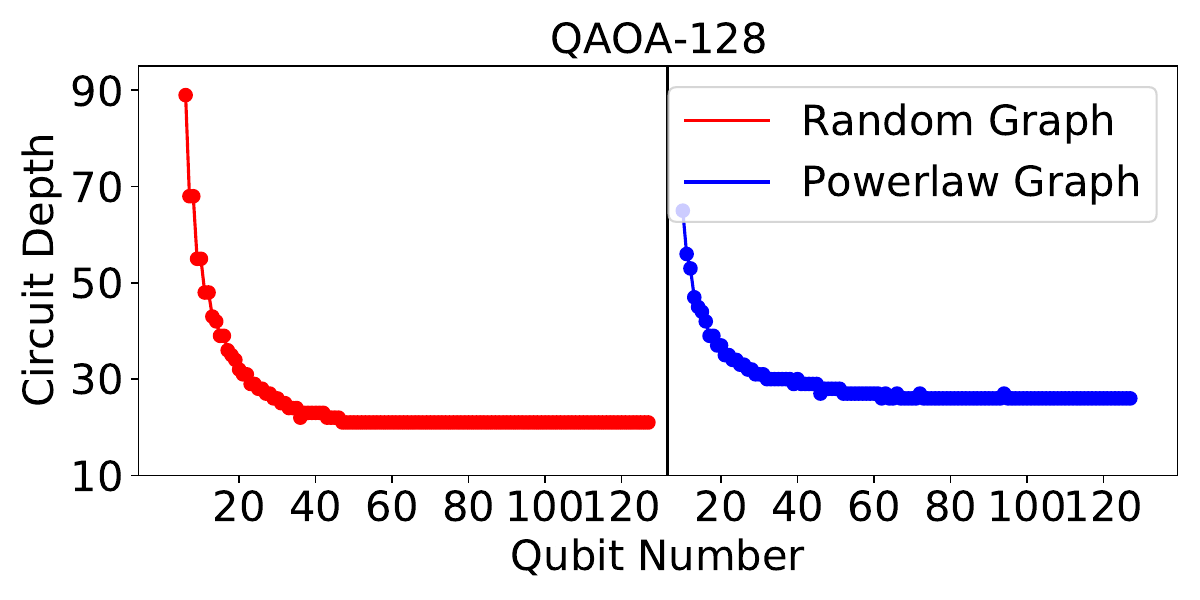}
         }
     
        \caption{Reuse vs depth for QAOA circuit (logical circuits)}
        \label{fig:QAOA_random_32}
\end{figure}

We also explore the trade-off between circuit depth and qubit usage in QAOA circuits. We did experiments on random and power-law graphs with the number of vertices of 16, 32, 64, and 128. Each of the graphs has a density of 30\%. We show results in Fig. \ref{fig:QAOA_random_32}. The result of graphs with 64 vertices are already shown in the motivation section, we only present results of 16, 32, and 128 qubits here. 

It turns out that QAOA circuit has more qubit reuse opportunities than regular quantum applications, especially for large cases. The minimum qubit usage of the power-law graph is closer to zero.  For both random graphs and power-law graphs, QAOA can save at least half of the qubits in the extreme case.  

Compared with the random graphs, the power-law graphs have more reuse, and a better tradeoff between depth and qubit number. This makes sense since the power-law graph contains more vertices with low degrees and the corresponding qubits have fewer gates on them. And the large degree node dominates the overall depth. This makes those qubits could be reused easily without sacrificing too much circuit depth.

\subsubsection{Tradeoff Analysis}
Since we generated different versions of circuits with respect to different qubit numbers for the same application, we can perform tradeoff analysis. We show the results in Table. \ref{tab:swap_best_q_usage}. 

We compare the version that has maximal reuse, the version that has minimal depth, and the baseline of Qiskit with highest optimization level. It shows that if targeting maximal reuse, then the circuit depth and duration will be affected adversely. If targeting minimal depth using the QS-CaQR version, we have moderate qubit saving. But for both versions of QS-CaQR, we are better than the baseline surprisingly for circuit depth/duration in a lot of cases. This demonstrates the usefulness of qubit reuse has extended beyond qubit saving.


\begin{table*}[t]
\centering
{
\centering
\caption{{QS-CaQR Version: The unit of circuit duration is \textsf{dt} -- system cycles. 1 dt is 0.22 nano-seconds.}}
\label{tab:swap_best_q_usage}
\resizebox{0.8\textwidth}{!}{
\begin{tabular}{|c|cccc|cccc|cccc|}
\hline
             & \multicolumn{4}{c|}{Baseline (No Reuse)}                                                     & \multicolumn{4}{c|}{Ours with Maximal Reuse}                                                   & \multicolumn{4}{c|}{Ours with Minimal Depth}                                                 \\ \hline
Benchmarks   & \multicolumn{1}{c|}{Qubit} & \multicolumn{1}{c|}{Depth} & \multicolumn{1}{c|}{Duration} & SWAP & \multicolumn{1}{c|}{Qubit} & \multicolumn{1}{c|}{Depth} & \multicolumn{1}{c|}{Duration} & SWAP & \multicolumn{1}{c|}{Qubit} & \multicolumn{1}{c|}{Depth} & \multicolumn{1}{c|}{Duration} & SWAP \\ \hline
RD-32        & \multicolumn{1}{c|}{4}    & \multicolumn{1}{c|}{28}    & \multicolumn{1}{c|}{88.6k}    & 9   & \multicolumn{1}{c|}{3}    & \multicolumn{1}{c|}{24}    & \multicolumn{1}{c|}{71.6K}    & 6     & \multicolumn{1}{c|}{3}    & \multicolumn{1}{c|}{24}    & \multicolumn{1}{c|}{71.6K}    & 6   \\ \hline
4mod5        & \multicolumn{1}{c|}{5}    & \multicolumn{1}{c|}{20}    & \multicolumn{1}{c|}{81.1K}    & 5   & \multicolumn{1}{c|}{4}    & \multicolumn{1}{c|}{18}    & \multicolumn{1}{c|}{61.5K}    & 3     & \multicolumn{1}{c|}{4}    & \multicolumn{1}{c|}{18}    & \multicolumn{1}{c|}{61.5K}    & 3   \\ \hline
Multiply\_13 & \multicolumn{1}{c|}{13}   & \multicolumn{1}{c|}{63}    & \multicolumn{1}{c|}{129K}     & 35  & \multicolumn{1}{c|}{8}    & \multicolumn{1}{c|}{58}    & \multicolumn{1}{c|}{145K}     & 17    & \multicolumn{1}{c|}{11}   & \multicolumn{1}{c|}{54}    & \multicolumn{1}{c|}{91.5K}    & 26  \\ \hline
System\_9    & \multicolumn{1}{c|}{12}   & \multicolumn{1}{c|}{279}   & \multicolumn{1}{c|}{431K}     & 95  & \multicolumn{1}{c|}{7}    & \multicolumn{1}{c|}{249}   & \multicolumn{1}{c|}{401K}     & 45    & \multicolumn{1}{c|}{6}    & \multicolumn{1}{c|}{230}   & \multicolumn{1}{c|}{314K}     & 39  \\ \hline
BV\_10       & \multicolumn{1}{c|}{10}   & \multicolumn{1}{c|}{30}    & \multicolumn{1}{c|}{92.3K}    & 18  & \multicolumn{1}{c|}{2}    & \multicolumn{1}{c|}{28}    & \multicolumn{1}{c|}{144K}     & 0     & \multicolumn{1}{c|}{7}    & \multicolumn{1}{c|}{23}    & \multicolumn{1}{c|}{72.4K}    & 5   \\ \hline
CC\_10       & \multicolumn{1}{c|}{10}   & \multicolumn{1}{c|}{15}    & \multicolumn{1}{c|}{88.7K}    & 9   & \multicolumn{1}{c|}{2}    & \multicolumn{1}{c|}{13}    & \multicolumn{1}{c|}{122K}     & 0     & \multicolumn{1}{c|}{5}    & \multicolumn{1}{c|}{10}    & \multicolumn{1}{c|}{72.6K}    & 1   \\ \hline
XOR\_5       & \multicolumn{1}{c|}{6}    & \multicolumn{1}{c|}{6}     & \multicolumn{1}{c|}{45.5K}    & 5   & \multicolumn{1}{c|}{2}    & \multicolumn{1}{c|}{7}     & \multicolumn{1}{c|}{42.7K}    & 0     & \multicolumn{1}{c|}{4}    & \multicolumn{1}{c|}{4}     & \multicolumn{1}{c|}{42.7K}    & 0   \\ \hline
QAOA5-0.3    & \multicolumn{1}{c|}{5}    & \multicolumn{1}{c|}{8}     & \multicolumn{1}{c|}{44.2K}    & 3   & \multicolumn{1}{c|}{3}    & \multicolumn{1}{c|}{5}     & \multicolumn{1}{c|}{34.5K}    & 1     & \multicolumn{1}{c|}{3}    & \multicolumn{1}{c|}{5}     & \multicolumn{1}{c|}{21.9K}    & 1   \\ \hline
QAOA10-0.3   & \multicolumn{1}{c|}{10}   & \multicolumn{1}{c|}{15}    & \multicolumn{1}{c|}{77K}      & 13  & \multicolumn{1}{c|}{7}    & \multicolumn{1}{c|}{11}    & \multicolumn{1}{c|}{65.5K}    & 4     & \multicolumn{1}{c|}{7}    & \multicolumn{1}{c|}{11}    & \multicolumn{1}{c|}{51.5K}    & 4   \\ \hline
QAOA15-0.3   & \multicolumn{1}{c|}{15}   & \multicolumn{1}{c|}{41}    & \multicolumn{1}{c|}{164K}     & 38  & \multicolumn{1}{c|}{4}    & \multicolumn{1}{c|}{48}    & \multicolumn{1}{c|}{207K}     & 0     & \multicolumn{1}{c|}{10}   & \multicolumn{1}{c|}{32}    & \multicolumn{1}{c|}{102K}     & 11  \\ \hline
QAOA20-0.3   & \multicolumn{1}{c|}{20}   & \multicolumn{1}{c|}{64}    & \multicolumn{1}{c|}{282K}     & 107 & \multicolumn{1}{c|}{4}    & \multicolumn{1}{c|}{72}    & \multicolumn{1}{c|}{509K}     & 0     & \multicolumn{1}{c|}{16}   & \multicolumn{1}{c|}{37}    & \multicolumn{1}{c|}{201K}     & 59  \\ \hline
QAOA25-0.3   & \multicolumn{1}{c|}{25}   & \multicolumn{1}{c|}{123}   & \multicolumn{1}{c|}{561K}     & 172 & \multicolumn{1}{c|}{5}    & \multicolumn{1}{c|}{133}   & \multicolumn{1}{c|}{820K}     & 0     & \multicolumn{1}{c|}{22}   & \multicolumn{1}{c|}{65}    & \multicolumn{1}{c|}{391K}     & 154 \\ \hline
\end{tabular}
}}
\end{table*}

\subsection{SR-CaQR Evaluation}
To evaluate the performance of SR-CaQR, we compare it with QS-CaQR.  
Firstly, we use SR-CaQR to compile the given circuit. Then for fairness, we use the version in QS-CaQR that has minimal SWAP number (we exhaust all possible qubit-saving count). Both experiments are conducted on IBM  Mumbai's architecture. 
We show results at Table \ref{tab:swap_reduction}.

\subsubsection{Regular Applications}
For all regular applications, SR-CaQR has the same or better SWAP gate count. For 4mod5 circuit, SR-CaQR minimizes the SWAP gate count to zero. For System\_9 circuit, SR-CaQR has 20.5\% of SWAP gate redcution.
\subsubsection{QAOA Applications}
For the QAOA applications, SR-CaQR has similar SWAP number compared with QS-CaQR for small applications since the near-optimal compilation is achieved by both solutions. For larger input graph size which has nodes larger than 15, the SR-CaQR uses fewer SWAPs and the duration time is also reduced. This demonstrates the usefulness of SR-CaQR.

\begin{table}[t]
{
\centering
\caption{{SR-CaQR (MIN-SWAP) v.s. QS-CaQR} }
\label{tab:swap_reduction}
\resizebox{0.45\textwidth}{!}{
\begin{tabular}{|c|cccc|cccc|}
\hline
             & \multicolumn{4}{c|}{QS-CaQR (MIN-SWAP)}                                                      & \multicolumn{4}{c|}{SR-CaQR}                                                                 \\ \hline
Benchmarks   & \multicolumn{1}{c|}{Qbt.} & \multicolumn{1}{c|}{Dpth.} & \multicolumn{1}{l|}{Drt.} & SWP & \multicolumn{1}{c|}{Qbt.} & \multicolumn{1}{c|}{Dpth.} & \multicolumn{1}{c|}{Drt.} & SWP \\ \hline
RD-32        & \multicolumn{1}{c|}{3}    & \multicolumn{1}{c|}{24}    & \multicolumn{1}{c|}{71.6K}    & 6   & \multicolumn{1}{c|}{3}    & \multicolumn{1}{c|}{24}    & \multicolumn{1}{c|}{71.6K}    & 6   \\ \hline
4mod5        & \multicolumn{1}{c|}{4}    & \multicolumn{1}{c|}{15}    & \multicolumn{1}{c|}{59.5K}    & 0   & \multicolumn{1}{c|}{4}    & \multicolumn{1}{c|}{15}    & \multicolumn{1}{c|}{59.5K}    & 0   \\ \hline
Multiply\_13 & \multicolumn{1}{c|}{11}   & \multicolumn{1}{c|}{54}    & \multicolumn{1}{c|}{91.5K}    & 26  & \multicolumn{1}{c|}{11}   & \multicolumn{1}{c|}{54}    & \multicolumn{1}{c|}{84.2K}    & 23  \\ \hline
System\_9    & \multicolumn{1}{c|}{6}    & \multicolumn{1}{c|}{230}   & \multicolumn{1}{c|}{314K}     & 39  & \multicolumn{1}{c|}{10}   & \multicolumn{1}{c|}{234}   & \multicolumn{1}{c|}{269K}     & 31  \\ \hline
BV\_10       & \multicolumn{1}{c|}{7}    & \multicolumn{1}{c|}{24}    & \multicolumn{1}{c|}{69.2K}    & 3   & \multicolumn{1}{c|}{7}    & \multicolumn{1}{c|}{24}    & \multicolumn{1}{c|}{69.2K}    & 3   \\ \hline
CC\_10       & \multicolumn{1}{c|}{6}    & \multicolumn{1}{c|}{9}     & \multicolumn{1}{c|}{70.2K}    & 2   & \multicolumn{1}{c|}{6}    & \multicolumn{1}{c|}{9}     & \multicolumn{1}{c|}{70.2K}    & 2   \\ \hline
XOR\_5       & \multicolumn{1}{c|}{4}    & \multicolumn{1}{c|}{0}     & \multicolumn{1}{c|}{42.7K}    & 0   & \multicolumn{1}{c|}{4}    & \multicolumn{1}{c|}{0}     & \multicolumn{1}{c|}{42.7K}    & 0   \\ \hline
QAOA5-0.3    & \multicolumn{1}{c|}{3}    & \multicolumn{1}{c|}{5}     & \multicolumn{1}{c|}{21.9K}    & 1   & \multicolumn{1}{c|}{3}    & \multicolumn{1}{c|}{5}     & \multicolumn{1}{c|}{21.9K}    & 1   \\ \hline
QAOA10-0.3   & \multicolumn{1}{c|}{7}    & \multicolumn{1}{c|}{11}    & \multicolumn{1}{c|}{51.5K}    & 4   & \multicolumn{1}{c|}{6}    & \multicolumn{1}{c|}{10}    & \multicolumn{1}{c|}{48.2K}    & 2   \\ \hline
QAOA15-0.3   & \multicolumn{1}{c|}{8}    & \multicolumn{1}{c|}{32}    & \multicolumn{1}{c|}{102K}     & 8   & \multicolumn{1}{c|}{8}    & \multicolumn{1}{c|}{27}    & \multicolumn{1}{c|}{98.2K}    & 5   \\ \hline
QAOA20-0.3   & \multicolumn{1}{c|}{15}   & \multicolumn{1}{c|}{37}    & \multicolumn{1}{c|}{201K}     & 52  & \multicolumn{1}{c|}{15}   & \multicolumn{1}{c|}{35}    & \multicolumn{1}{c|}{195K}     & 51  \\ \hline
QAOA25-0.3   & \multicolumn{1}{c|}{20}   & \multicolumn{1}{c|}{65}    & \multicolumn{1}{c|}{391K}     & 124 & \multicolumn{1}{c|}{20}   & \multicolumn{1}{c|}{63}    & \multicolumn{1}{c|}{360K}     & 114 \\ \hline
\end{tabular}
}}
\end{table}

\subsection{Real machine experiment}
We run BV\_5, BV\_10, Multiply\_13, CC\_13 circuit, QAOA 10-0.3, QAOA 10-0.5 circuit at IBM Mumbai device. This is the one machine supporting dynamic circuit. Not all IBM machine support that. For regular circuits, we use TVD to evaluate the output distribution. The results are shown in Table \ref{tab:TVD}. A takeaway is that our SR-CaQR has improved for all the benchmarks listed here. Since the current real machine is still in an early stage and the mid-measurement pulse is not stable, so we hope to see results for all these benchmarks once the machine supporting dynamic circuit becomes more mature.  

\begin{figure}
    \centering
    \includegraphics[width=0.4\textwidth]{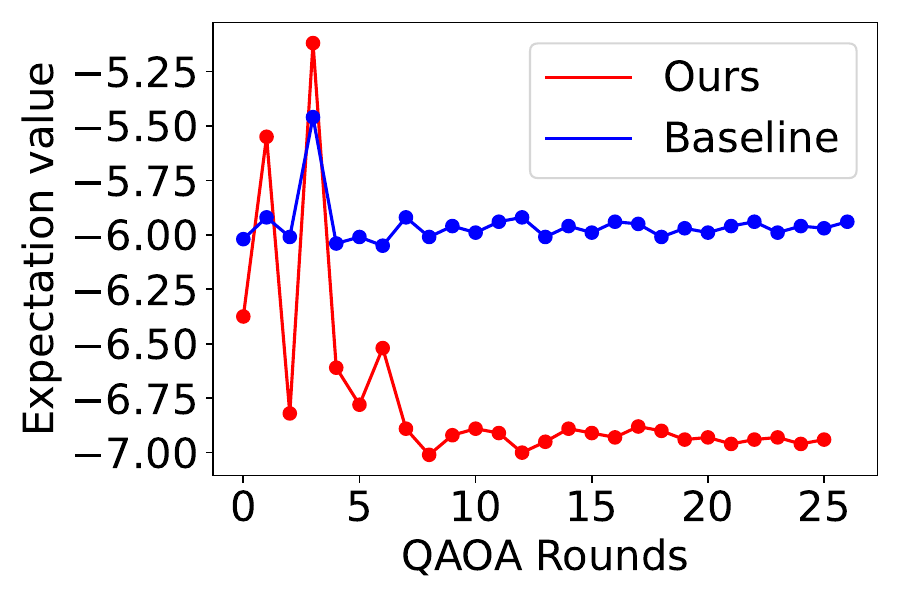}
    \caption{The end to end result of the QAOA reuse experiments for QAOA-10 with density 30\% }
    \label{fig:real-exp-qaoa10-0.3}
\end{figure}
\begin{figure}
    \centering
        \includegraphics[width=0.4\textwidth]{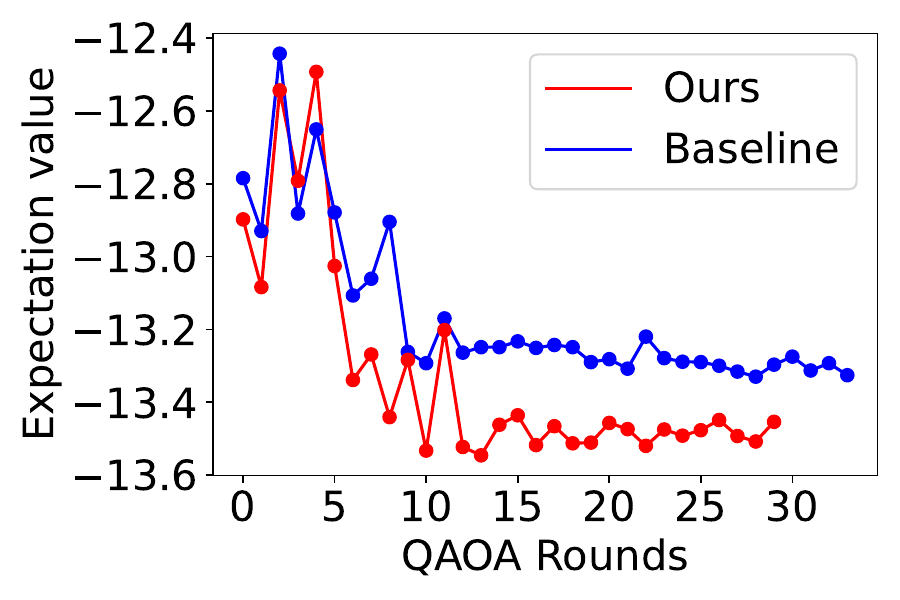}
    \caption{The end to end result of the QAOA reuse experiments for QAOA-10 with density 50\%.}
    \label{fig:real-exp-qaoa10-0.5}
\end{figure}

For QAOA circuits, the results are shown in Fig. \ref{fig:real-exp-qaoa10-0.3} and Fig. \ref{fig:real-exp-qaoa10-0.5}. For both figures, the x-axis stands for the round number of the parameter optimization by classical machine learning optimizer called COBYLA. The y-axis is the negation of the expected value of the max-cut value. The smaller is better.   The red curve is the result of SR-CaQR with 6 qubits and the blue curve is the result of the circuit without qubit reuse. For both experiments, SR-CaQR circuits achieve better max-cut values and converge faster.


\begin{table}[]
\centering
\caption{TVD results for BV, Multiply, CC circuit }
\label{tab:TVD}
\resizebox{0.35\textwidth}{!}{
\begin{tabular}{|c|c|c|}
\hline
\multicolumn{1}{|l|}{Benchmarks} & TVD (Baseline) & TVD (SR-CaQR) \\ \hline
Multiply\_13                     & 0.76           & 0.61          \\ \hline
BV\_10                           & 0.64           & 0.48          \\ \hline
CC\_10                           & 0.61           & 0.44          \\ \hline
\end{tabular}
}
\end{table}


\section{Related Work}
\label{sec:rel}

Generic compilers, such as \cite{zhang+:asplos21, Li+:ASPLOS19, murali+:asplos19,siraichi+:oopsla19,Siraichi+:CGO18,Zulehner+:DATE18,Zulehner+:ICRC17, tan+:iccad20, tannu+:asplos19, Wille+:DAC19, Shafaei+:ASPDAC14} compile the given quantum circuit with the consideration of circuit depth, gate count, and qubits variability. After the applications with gate commutativity such as QAOA \cite{QAOA:farhi2014quantum,QAOA:farhi2017quantum,QAOA:farhi2021quantum} and VQE \cite{vqe:Peruzzo2014} gained more attention, domain-specific compilers exploit the commutativity feature. Some new compilers exploiting the such feature are proposed \cite{lao+:isca20,alam+:micro20,li+:asplos22}. But none of those are aware of the opportunities in qubit reuse.

Paler \etal \cite{paler+:physreva16}  proposed a method, named wire recycling, for quantum circuit synthesis with the same number of qubit but less wire. They construct a causal graph from the original circuit and utilize the graph search algorithm to find such qubit wire recycling opportunities. Even the wire recycling only applies to the ancilla qubit in the quantum reversible circuit, their method could be adapted to find the qubit reuse opportunity in QAOA. In addition, they also provide a table of reversible circuits that proves the qubit reuse is pragmatic. But in their cases, ancilla qubits are already pre-defined. 

DeCross \etal \cite{decross+:arxiv22midcircuit} proposed a reuse compilation on all-to-all connectivity ion-trap machines. This is a simultaneous study. We notice their work when we are about to submit our work on Arxiv. They use SAT to find the maximum usage for small circuit and heuristic method for large circuit. The main difference here is that we consider the swaps insertions while they did not. Our real experiments are done on a superconducting quantum computer at IBM while they focus on ion-trap machine.

SQAURE\cite{Ding+:ISCA20} is a qubit reuse compilation framework built upon the uncomputation strategy. It has a locality-aware allocation strategy to decide which ancilla qubit to be reused to reduce the circuit communication overhead. It also has a cost-cffective reclamation strategy to decide where the uncomputation should be applied. However, this compilation framework leveraging the uncomputation can only be applied to reversible arithmetic circuits and reclaim ancilla qubits. In addition, the uncomputation strategy  inevitably introduces more gates to reset ancilla qubit back to state |0>. As a result, it makes compiled circuit have a longer circuit depth and larger accumulated gate errors. Note that the reclamation through un-computation is limited to ancilla qubit only. In our case, we still need the outcome result of the qubits we measure, before reusing them. Ancilla qubits do not need to be measured.

Govia \etal \cite{Govia:arxiv} proposed a randomized benchmark suite for mid-circuit measurements. It can be used to test the impact of mid-circuit measurement.  This work is meaningful in device characterization and is complementary to our work.

\section{Conclusion}
\label{sec:conclusion}

With supported mid-circuit hardware measurement, we
can improve circuit efficacy and fidelity from three aspects:
(a) reduced qubit usage, (b) reduced swap insertion, and (c)
improved estimated success probability. We demonstrate this
using real-world applications Bernstein Verizani on real hardware and show that circuit resource usage can be improved by
60\%, and circuit fidelity can be improved by 15\%. We design
a compiler-assisted tool that can find and exploit the tradeoff
between qubit reuse, fidelity, gate count, and circuit duration.

\bibliographystyle{plain}
\bibliography{references, allrefs, newref}

\end{document}